\documentclass[%
 reprint,
 amsmath,amssymb,
 aps,
]{revtex4-2}

\usepackage{graphicx}% Include figure files
\usepackage{dcolumn}% Align table columns on decimal point
\usepackage{bm}% bold math
\usepackage{amsmath}

\usepackage{diagbox}
\usepackage{pifont}
\usepackage{multirow}
\usepackage{ctable}
\begin{document}

\preprint{APS/123-QED}

\title{Compression-Complexity with Ordinal Patterns for Robust Causal Inference in Irregularly-Sampled Time Series}% Force line breaks with \\
%Robust Causal Inference using Permutation Patterns and Compression-Complexity: Applications on Challenging Climate Data
%Ordinal Patterns Combined with Compression-Complexity for Causal Inference on Challenging Data
%Ordinal Patterns for Causal Inference using Datasets with Missing and Finite Samples
%Ordinal pattern based compression complexity causality for missing and finite data
%Ordinal patterns and compression-complexity based causal inference for missing and finite data
% Permutation/Ordinal patterns combined with Compression-Complexity Causality for causal inference on challenging data

\author{Aditi Kathpalia}
 \email{kathpalia@cs.cas.cz}
\author{Pouya Manshour}%
 \author{Milan Palu\v{s}}%
 \email{mp@cs.cas.cz}
\affiliation{%
 Department of Complex Systems, \\
 Institute of Computer Science of the Czech Academy of Sciences, \\
 Prague, Czech Republic
}%


%  \homepage{http://www.Second.institution.edu/~Charlie.Author}
% \affiliation{
%  Second institution and/or address\\
%  This line break forced% with \\
% }%
% \affiliation{
%  Third institution, the second for Charlie Author
% \affiliation{%
%  Authors' institution and/or address\\
%  This line break forced with \textbackslash\textbackslash
% }%

% \collaboration{CLEO Collaboration}%\noaffiliation

\date{\today}% It is always \today, today,

\begin{abstract}
Distinguishing cause from effect is a scientific challenge resisting solutions from mathematics, statistics, information theory and computer science. \emph{Compression-Complexity Causality (CCC)} is a recently proposed \emph{interventional} measure of causality, inspired by Wiener-Granger’s idea. It estimates causality based on change in dynamical compression-complexity (or compressibility) of the effect variable, given the cause variable. CCC works with minimal assumptions on given data and is robust to irregular-sampling, missing-data and finite-length effects. However, it only works for one-dimensional time series. We propose an ordinal pattern symbolization scheme to encode multidimensional patterns into one-dimensional symbolic sequences, and thus introduce the \emph{Permutation CCC (PCCC)}, which retains all advantages of the original CCC and can be applied to data from multidimensional systems with potentially hidden variables. PCCC is tested on numerical simulations and applied to paleoclimate data characterized by irregular and uncertain sampling and limited numbers of samples.
%\begin{description}
%\end{description}
\end{abstract}

\maketitle

%\tableofcontents

\section{\label{sec:level1}Introduction}

Unraveling systems' dynamics from the analysis of observed data is one of the fundamental goals of many areas of natural and social sciences. In this respect, detecting the direction of interactions or inferring causal relationships among observables is of particular importance that can improve our ability to better understand the underlying dynamics and to predict or even control such complex systems ~\cite{pearl2018book, kathpalia2021measuring}.

Around sixty years after the pioneering work of Wiener and Granger~\cite{wiener, granger} on quantifying linear `causality' from observations, it has been widely applied not only in economics~\cite{geweke1984inference, hiemstra1994testing, chiou2008economic}, for which it was first introduced, but also in various fields of natural sciences, from neurosciences~\cite{seth2015granger} to Earth sciences~\cite{mosedale2006granger, tirabassi2015study, runge2019inferring}. A number of attempts have been made to generalize Granger Causality (GC) to nonlinear cases, using, e.g., an estimator based on correlation integral~\cite{hiemstra1994testing}, a non-parametric regression approach~\cite{bell1996non}, local linear predictors~\cite{chen2004analyzing}, mutual nearest neighbors~\cite{schiff1996detecting, le1999nonlinear}, kernel estimators~\cite{marinazzo}, to state a few. Several other causality methods based on the GC principle such as Partial Directed Coherence~\cite{baccala2001partial}, Direct Transfer Function~\cite{kaminski2001evaluating} and Modified Direct Transfer Function~\cite{korzeniewska2003determination} have also been proposed.

Information theory has proved itself as a powerful approach into causal inference. In this respect, Schreiber proposed a method for measuring information transfer among observables~\cite{schreiber}, known as \textit{Transfer Entropy} (TE), which is based on Kullback-Leibler distance between transition probabilities. Palu\v{s} et al.~\cite{paluvs2001synchronization} introduced a causality measure based on mutual information, called \textit{Conditional Mutual Information} (CMI). CMI has been shown to be equivalent to TE~\cite{paluvs2007directionality}. These tools have been applied in various research studies and have shown their power in extracting causal relationships between different systems~\cite{vicente2011transfer, bauer2007finding, dimpfl2013using, paluvs2014multiscale, jajcay2018synchronization}.

We usually work with time series $x(t)$ and $y(t)$ as realizations of $m$ and $n$ dimensional dynamical systems, $X(t)$ and $Y(t)$ respectively, evolving in measurable spaces. It means that $x(t)$ and $y(t)$ can be considered as the components of these $m$ and $n$ dimensional vectors. In many cases, only one possible dimension of the phase space is observable, recordings or knowledge of variables which may have indirect effects or play as mediators in the causal interactions between observables may not be available. In this respect, phase-space reconstruction is a common useful approach introduced by Takens~\cite{taken}, which reconstructs the dynamics of the entire system (including other 
unknown/unmeasurable variables) using time-delay embedding vectors, as follows: the manifold of an $m$ dimensional state vector $X$ can be reconstructed as $X(t)=\{x(t),x(t-\eta),...,x(t-(m-1)\eta)\}$. Here, $\eta$ is the embedding delay, and can be obtained using the embedding construction procedure based on the first minimum of the mutual information~\cite{fraser1986independent}. Some causality estimators have applied this phase-space reconstruction procedure to improve their causal inference power, such as high dimension CMI~\cite{paluvs2014multiscale}  and TE~\cite{wibral2013measuring}. Other causality measures, such as, Convergent Cross Mapping (CCM)~\cite{sugihara}, Topological Causality~\cite{harnack2017topological}, Predictability Improvement~\cite{krakovska2016testing}, are based directly on the reconstruction of dynamical systems.

Vast amounts of data available in the recent years have pushed some of the above discussed GC extensions, information and phase-space reconstruction based approaches forward as they rely on joint probability density estimations, stationarity, markovianity, topological or linear modeling. However, still, many temporal observations made in various domains such as climatology~\cite{barrios2018alternative, anderson2018accounting}, finance~\cite{dicesare2006imputation, john2019imputation} and sociology~\cite{gyimah2001missing} are often short in length, have missing samples or are irregularly sampled. A significant challenge arises when we attempt to apply causality measures in such situations~\cite{runge2019inferring}. For instance, CMI or TE fail when applied to time series which are undersampled or have missing samples~\cite{kulp2009application, smirnov2012spurious, kathpalia2019data} and also in case of time series with short lengths~\cite{kathpalia2019data}. CCM and kernel based non-linear GC also show poor performance even in the case of few missing samples in bivariate simulated data~\cite{kathpalia2021theoretical}.

Kathpalia and Nagaraj recently introduced a causality measure, called Compression-Complexity Causality (CCC), which employs `complexity' estimated using lossless data-compression algorithms for the purpose of causality estimation. It has been shown to have the strength to work well in case of missing samples in data for bivariate systems of coupled autoregressive and tent map processes. This has been shown to be the case for samples which are missing in the two coupled time series either in a synchronous or asynchronous manner~\cite{kathpalia2019data}. Also, it gives good performance for time series with short lengths~\cite{kathpalia2019data, kathpalia2021theoretical}. These strengths of CCC arise from its formulation as an \emph{interventional} causality measure based on the evolution of dynamical patterns in time series, independence from joint probability density functions, making minimal assumptions on the data and use of lossless compression based complexity approaches which in turn show robust performance on short and noisy time series~\cite{kathpalia2019data, nagaraj2}.  However, as discussed in~\cite{kathpalia2021theoretical}, a direct multidimensional extension of CCC is not as straightforward and so a measure of \emph{effective CCC} has been formulated and used on multidimensional systems of coupled autoregressive processes with limited number of variables.

On the other hand, a method for symbolization of phase-space reconstructed (embedded) processes has been used to improve the ability of info-theoretic causality measures for noisy data, such as \emph{symbolic transfer entropy} \cite{staniek2008symbolic, staniek2009symbolic}, \emph{partial symbolic transfer entropy}~\cite{kugiumtzis2013partial, papana2013simulation}, \emph{permutation conditional mutual information} (PCMI) \cite{li2010estimating} and \emph{multidimensional PCMI}~\cite{wen2019estimating}. The symbolization technique used in these works is based on the Bandt and Pompe scheme for estimation of Permutation Entropy~\cite{bandt2002permutation}, and often referred to as \emph{permutation} or \emph{ordinal patterns} coding. The scheme labels the embedded values of time-series at each time point in ascending order of their magnitude. Symbols are then assigned at each time point depending on the ordering of values (or the labelling sequence) at that point. Ordinal patterns have been used extensively in the analysis and prediction of chaotic dynamical systems and also shown to be robust in applications to real world time series. By construction, this technique ignores the amplitude information and thus decreases the effect of high fluctuations in data on the obtained causal inference~\cite{fadlallah2013weighted}. Other benefits of permutation patterns are: they naturally emerge from the time series and so the method is almost parameter-free; are invariant to monotonic transformations of the values; keep account of the causal order of temporal values and the procedure is computationally inexpensive~\cite{amigo2010permutation, zanin2012permutation, keller2014ordinal, zanin2021ordinal}. Ordinal partition has been shown to have the generating property under specific conditions, implying topological conjugacy between phase space of dynamical systems and their ordinal symbolic dynamics~\cite{mccullough2015time}. Further, permutation entropy for certain sets of systems has been shown to have a theoretical relationship to the system's Lyapunov exponents and Kolmogorov Sinai Entropy~\cite{bandt2002entropy, amigo2005permutation}. Because of all these beneficial properties of permutation patterns, it is no wonder that the development of symbolic TE or PCMI helped to make them more robust, giving better performance in the case of noisy measurements, simplifying the process of parameter selection and making less demands on the data.
%If the embedding dimension of the system is m, there will be m factorial unique values in the sequence.

In this work, we propose the use of CCC approach with reconstructed dynamical systems which are symbolized using ordinal patterns. The combination of strengths of CCC and ordinal patterns, not only makes CCC applicable to dynamical systems with multidimensional variables, but we also observe that the proposed \emph{Permutation CCC} (PCCC) measure gives great performance on datasets with very short lengths and high levels of missing samples. The performance of PCCC is compared with that of PCMI (which is identical to symbolic TE), bivariate CCC and CMI on simulated dynamical systems data. PCCC outperforms the existing approaches and its estimates are found to be robust for short length time series, and high levels of missing data points.

This development for the first time opens up avenues for the use of causality estimation tool on real world datasets from climate and paleoclimate science, finance and other fields where there is prevalence of data
with irregular and/or uncertain sampling times. 
To determine the major drivers of climate is the need of the hour as climate change poses a big challenge to humankind and our planet Earth~\cite{solomon2007climate}. Different studies have employed either correlation/coherence, causality methods or modelling approaches to study the interaction between climatic processes. The results produced by different studies are different and sometimes contradictory, presenting an ambiguous situation. We apply PCCC to analyse the causal relationship between the following sets of climatic processes: greenhouse gas concentrations -- atmospheric temperature, El-Niño Southern Oscillation -- South Asian monsoon and North Atlantic Oscillation -- European temperatures at different time-scales and compare its performance with bivariate CCC, bivariate and multidimensional CMI, and PCMI. The time series available for most of these processes are short in length and sometimes have missing samples and (or) are sampled in irregular intervals of time. We expect our estimates to be reliable and to be helpful to resolve the ambiguity presented by existing studies.

\section{\label{sec:level2}Results}

\textbf{Simulation Experiments}:
Time series data from a pair of unidirectionally coupled R{\"o}ssler systems were generated as per the following equations:
\begin{eqnarray}\label{aros1}
\mathaccent95 x_{1} &= &- \omega_{1} y_{1} - z_{1}, \nonumber \\
\mathaccent95 y_{1} &= &\omega_{1} x_{1} + a_1 \; y_{1}, \\
\mathaccent95 z_{1} &= &b_1 + z_{1} (x_{1} - c_1), \nonumber
\end{eqnarray}
for the autonomous or master system, and
\begin{eqnarray}\label{aros2}
\mathaccent95 x_{2} &= &- \omega_{2} y_{2} - z_{2} + \epsilon
(x_{1} - x_{2}), \nonumber \\
\mathaccent95 y_{2} &= &\omega_{2} x_{2} + a_2 \; y_{2}, \\
\mathaccent95 z_{2} &= &b_2 + z_{2} (x_{2} - c_2), \nonumber
\end{eqnarray}
\noindent for the response or slave system. Parameters were set as: $a_1 = a_2 = 0.15$, $b_1 = b_2 = 0.2$, $c_1 = c_2 = 10.0$, and frequencies set as: $\omega_{1} = 1.015$ and $\omega_{2} = 0.985$. The coupling parameter, $\epsilon$, was fixed to $0.09$. The data were generated by numerical integration based on the adaptive Bulirsch-Stoer method \cite{press1987numerical} using a sampling interval of 0.314 for both the master and slave systems. This procedure gives 17 -- 21 samples per one period. 100 realizations of these systems were simulated and initial 5000 transients were removed before using the data for testing experiments.

As can be seen from the equations, there is a coupling between $x_1$ and $x_2$, with $x_1$ influencing $x_2$. The analysis of the causal influence between the two systems was done using the causality estimation measures: bivariate or scalar CCC, CMI, PCCC and PCMI for the cases outlined in the following paragraphs. The estimation procedure for each of the methods is described in the `Methods' section. The values of parameters used for each of the methods are also given in the `Methods' section (Table~\ref{table_params}).

\begin{figure}
\centering
\includegraphics[width=\columnwidth]{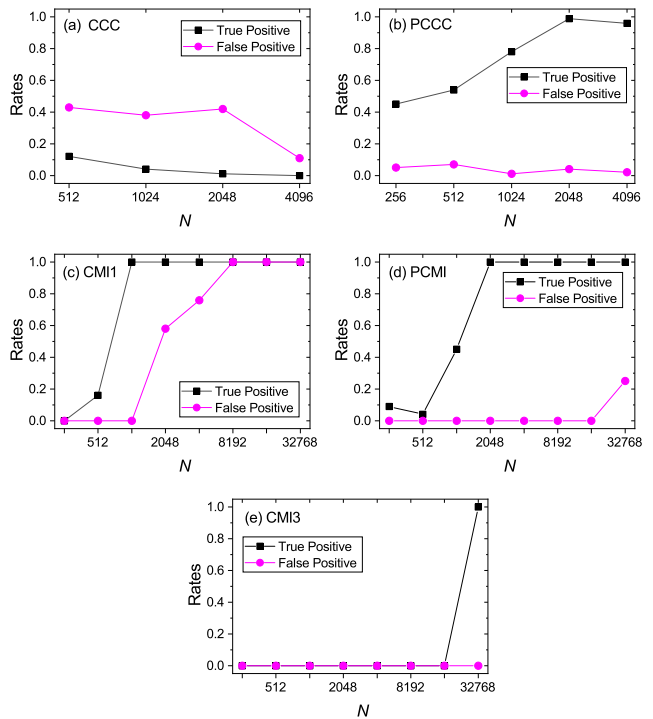}% Here is how to import EPS art
\caption{\label{fig_sim_len} Specificity and sensitivity of methods with varying length. True positive rate (or rate of significant causality estimated from $x_1 \rightarrow x_2$) and false positive rate (or rate of significant causality estimated from $x_2 \rightarrow x_1$), using measures (a) scalar CCC (CCC), (b) permutation CCC (PCCC), (c) scalar CMI (CMI1), (d) permutation CMI (PCMI) and (e) three-dimensional CMI (CMI3), as the length of time series, $N$, is varied.}
\end{figure}

\emph{Finite length data:} The length of time series, $N$, of $x_1$ and $x_2$  taken from coupled R{\"o}ssler systems was varied as shown in Fig.~\ref{fig_sim_len}. The estimation for CMI and PCMI is done up to a higher value of length as CMI did not give optimal performance until the length became 32,768 samples. Fig.~\ref{fig_sim_len}(c) shows scalar (simple bivariate) CMI or one-dimensional CMI (CMI1) between $x_1$ and $x_2$ (see Palu\v{s} and Vejmelka~\cite{paluvs2007directionality}). This method has high sensitivity but suffers from low specificity. This problem is solved by using conditional CMI or three-dimensional CMI (CMI3), where the information from other variables ($y_1, z_1, y_2, z_2$) is incorporated in the estimation. Its performance is depicted in Fig.~\ref{fig_sim_len}(e). However, it requires larger length of time series for optimal performance. Fig.~\ref{fig_sim_len}(a) shows the performance of scalar (or simple bivariate) CCC, which is equivalent to the CMI1 case, considering dimensionality. Figs.~\ref{fig_sim_len}(b) and~\ref{fig_sim_len}(d) show the performance of PCCC and PCMI respectively. For each length level, all 100 realizations of coupled systems were considered and 100 surrogates generated for each realization in order to perform significance analysis of causality estimated (in both directions) from each realization of coupled processes. These surrogates were generated for both the processes using the Amplitude Adjusted Fourier Transform method~\cite{theiler1992testing} and significance testing done using a standard one-sided z-test with p-value set to 0.05 (this was justified as the distributions of surrogates for CCC and CMI methods implemented were found to be Gaussian). Based on this significance analysis, true positive rate (TPR) and false positive rate (FPR) were computed at each length level. A true positive is counted for a particular realization of coupled systems when causality estimated from $x_1$ to $x_2$ is found to be significant and a false positive is counted when causality estimated from $x_2$ to $x_1$ is found to be significant.

As it can be seen from the plots, direct application of scalar CCC completely fails on multidimensional dynamical systems data, yielding low true positives and high false positives. Hence the method displays poor sensitivity as well as specificity. CMI1 also shows poor performance, yielding high false positives. CMI3, which is appropriate to be applied for multi-dimensional data, only begins to give good performance when the length of time series is taken to be greater than 32,768 samples. On the other hand, PCCC begins to give high true positives and low false positives, as the length of time series is increased to 1024 time points, with TPR and FPR reaching almost 1 and 0 respectively as length is increased to 2048 time points. The use of permutation patterns also improves the performance of CMI3 for short length data as it can be seen that PCMI begins to show a TPR of 1 and FPR of 0 for length of time series equal to 2048 time points.

We did further experiments with simulated R{\"o}ssler data by varying the amount of noise and missing samples in the data. For these cases, performance of PCCC and PCMI alone were evaluated because it can be seen from the `varying length' experiments that scalar CCC and CMI1 do not work for multidimensional dynamical systems data and CMI3 does not perform well for short length data.

\begin{figure}
\centering
\includegraphics[width=\columnwidth]{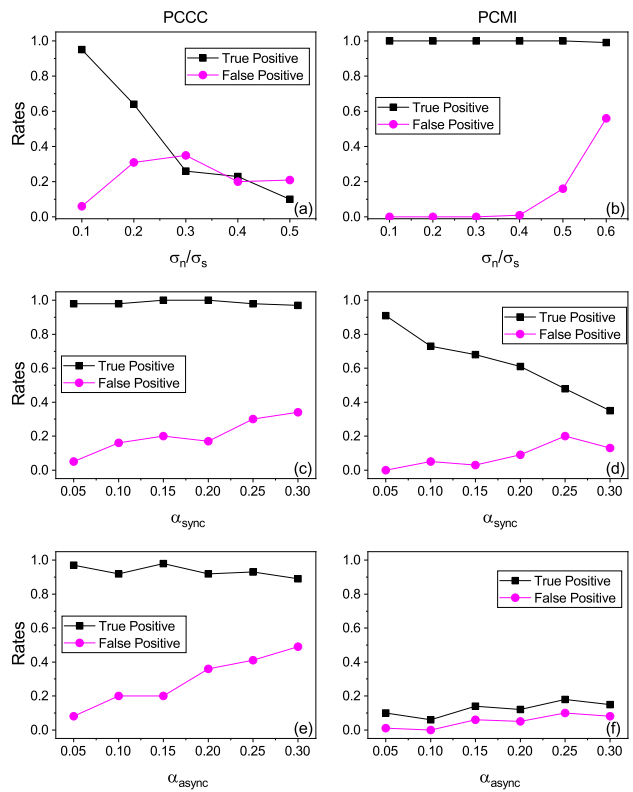}% Here is how to import EPS art
\caption{\label{fig_sim_noise_spar} Specificity and sensitivity of methods with varying noise and sparsity. True positive rate (or rate of significant causality estimated from $x_1 \rightarrow x_2$) and false positive rate (or rate of significant causality estimated from $x_2 \rightarrow x_1$), using measures permutation CCC (PCCC) (left column) and permutation CMI (PCMI) (right column) as the level of noise: (a) and (b); level of synchronous sparsity: (c) and (d); and asynchronous sparsity: (e) and (f), are varied.}
\end{figure}

\emph{Noisy data:} White Gaussian noise was added to the simulated R{\"o}ssler data. The amount of noise added to the data was relative to the standard deviation of the data. The noise standard deviation ($\sigma_n$), is expressed as a percentage of the standard deviation of the original data ($\sigma_s$). For example, $20\%$ of noise means $\sigma_n= 0.2\sigma_s$, and $100\%$ of noise means $\sigma_n= \sigma_s$. The length of time series taken for this experiment was fixed to 2048. For each realization of noisy data as well, 100 surrogate time series were generated and significance testing performed as before using the Amplitude Adjusted Fourier Transform method and z-test respectively. Figs.~\ref{fig_sim_noise_spar}(a) and~\ref{fig_sim_noise_spar}(b) show the results for varying noise in the data for the measures PCCC and PCMI respectively.

It can be seen that PCCC performs well for low levels of noise, up to $10\%$, but at higher levels of noise, its performance begins to deteriorate. PCMI, on the other hand, shows high TPR and low FPR even as the noise level is increased to $50\%$.

\emph{Sparse data:} We refer to time-series with missing samples as sparse data. Sparsity or non-uniformly missing samples were introduced in the data in two ways: (1) Synchronous sparsity and (2) Asynchronous sparsity. In case of (1), samples were missing from both $x_1$ and $x_2$ at randomly chosen time indices and this set of time indices was the same for both $x_1$ and $x_2$. In case of (2), samples were missing from both $x_1$ and $x_2$ based on two different sets of randomly chosen time indices, that is, the time indices of missing samples were different for $x_1$ and $x_2$. The amount of synchronous/ asynchronous sparsity is expressed in terms of percentage of missing samples relative to the original length of time series taken. $\alpha_{sync}$ and $\alpha_{async}$ refer to the level of missing samples for the cases of synchronous and asynchronous sparsity respectively, and are given by $m/N$, where $m$ is the number of missing samples and $N$ is the original length of time series. $N$ was fixed to 2048. The length of time series became shorter as the percentage of missing samples were increased. Causality estimation measures were applied to the data without any knowledge of whether any samples were missing or the time stamps at which the samples were missing. Surrogate data generation for each realization in this case was not done post the introduction of missing samples but prior to that, using the original length time series. Sparsity was then introduced in the surrogate time series in a manner similar to that for original time series.

Figs.~\ref{fig_sim_noise_spar}(c) and~\ref{fig_sim_noise_spar}(d) show the results obtained using PCCC and PCMI respectively for synchronous sparsity. Figs.~\ref{fig_sim_noise_spar}(e) and~\ref{fig_sim_noise_spar}(f) show the same for asynchronous sparsity. It can be seen that PCCC is robust to the introduction of missing samples, showing high TPR and low FPR. FPR begins to be greater than 0.2 only when the level of synchronous sparsity is increased to $25\%$ and asynchronous sparsity is increased to $20\%$. PCMI is robust to low levels of synchronous sparsity but deteriorates beyond $5\%$ of missing samples, giving low true positives. It performs very poorly even with low levels of asynchronous sparsity.

\textbf{Real Data Analysis}:
As discussed in the Introduction, a number of climate datasets are either sampled at irregular intervals, have missing samples, are sampled after long intervals of time or have a combination of two or more of these issues. In addition, their temporal recordings available are short in length. We apply the proposed method, PCCC, to some such datasets described below. We also compare the results obtained with existing measures: scalar CCC, scalar CMI and PCMI.

\emph{Millenial scale CO\textsubscript{2}-temperature recordings:} Mills et al. have compiled independent estimates of global average surface temperature and atmospheric CO\textsubscript{2} concentration for the Phanerozoic eon. These paleoclimate proxy records span the last 424 million years~\cite{mills2019modelling} and have been used and made available in the study by Wong et al.~\cite{wong2021tighter}. One data point for both CO\textsubscript{2} and temperature recordings were available for each million year period and was used in our analysis to check for causal interaction between between the two.

\emph{CO\textsubscript{2}, CH\textsubscript{4} and temperature recordings over the last 800,000 years:}
In~\cite{past2016interglacials}, Past Interglacials Working Group of PAGES has made available proxy records of atmospheric CO\textsubscript{2}, CH\textsubscript{4} and deepwater temperatures over the last 800 ka (1 ka= 1000 years). Each of these time series were reconstructed by separate studies and so the recordings available are non-synchronous and also irregularly sampled for each variable. Further, some data points are missing in the temperature time-series. Roughly, single data point is available for each ka for each of the three variables. CO\textsubscript{2} proxy data are based on antarctic ice core composites. This was first reported in~\cite{luthi2008high} and the revised values made available in a study by Bereiter et al.~\cite{bereiter2015revision}. Reconstructed atmospheric CH\textsubscript{4} concentrations, also based on ice cores, were as reported in~\cite{loulergue2008orbital} (on the AICC2012 age scale~\cite{bazin2013optimized}). Deepwater temperature recordings obtained using shallow-infaunal benthic foraminifera (Mg/Ca ratios) that became available from Ocean Drilling Program (ODP) site 1123 on the Chatham Rise, east of New Zealand were reported in~\cite{elderfield2012evolution}.

Causal influence was checked between CO\textsubscript{2}-temperature and separately between CH\textsubscript{4}-temperature. CO\textsubscript{2} and CH\textsubscript{4} data are taken beginning from the 6.5\textsuperscript{th} ka on the AICC2012 scale and temperature data are taken beginning from the 7\textsuperscript{th} ka. Since the number of data points available for temperature are 792, CO\textsubscript{2}-temperature analysis was done based on these 792 samples and as the number of samples of CH\textsubscript{4} is limited to 756 beginning from the 6.5\textsuperscript{th} ka, CH\textsubscript{4}-temperature analysis was done using these 756 data points.

\emph{Monthly CO\textsubscript{2}-temperature dataset:} Monthly mean CO\textsubscript{2} data constructed from mean daily CO\textsubscript{2} values as well as Northern Hemisphere's combined land and ocean temperature anomalies for the monthly timescale are available open source on the National Oceanic and Atmospheric
Administration (NOAA) website. The CO\textsubscript{2} measurements were made at the Mauna Loa Observatory, Hawaii. A part of the CO\textsubscript{2} dataset (March 1958-April 1974) were originally obtained by C. David Keeling of the Scripps Institution of Oceanography and are available on the Scripps website. NOAA started its own CO\textsubscript{2} measurements starting May 1974. The temperature anomaly dataset is constructed from the Global Historical Climatology Network-Monthly data set~\cite{lawrimore2011overview} and International Comprehensive Ocean-Atmosphere Data Set, also available on the NOAA website. These data from March, 1958 to June 2021 (with 760 data points) were used to check for the causal influence between CO\textsubscript{2} and temperature on the recent timescale. Both time series were differenced using consecutive values as they were highly non-stationary.

\emph{Yearly ENSO-SASM dataset:} 1,100 Year El Niño/Southern Oscillation (ENSO) Index Reconstruction dataset, made available open source on NOAA website and originally published in~\cite{li2011interdecadal} was used in this study. South Asian Summer Monsoon (SASM) Index 1100 Year Reconstruction dataset, also available open source on the NOAA website and originally published in~\cite{shi2014tree}, was the second variable used here. The aim of our study was to check the causal dependence between these two sets of recordings taken from the year 900 AD to 2000 AD (with one data point being available for each year).

\emph{Monthly NINO-Indian monsoon dataset:} Monthly NINO 3.4 SST Index recordings from the year 1870 to 2021 are available open source on the NOAA website. Its details are published in~\cite{rayner2003global}. All India monthly rainfall dataset from 1871-2016, available on the official website of World Meteorological Organization and originally acquired from `Indian Institute of Tropical Meteorology', was used for analysis. These recordings are in the units of mm/month. Causal influence was checked between these two recordings using 1752 data points, ranging from the month January, 1871 to December, 2016.

\emph{Monthly NAO-temperature recordings:} Reconstructed monthly North Atlantic Oscillation (NAO) index recordings from December 1658 to July 2001 are available open source on the NOAA website. The reconstructions from December 1658 to November 1900 are taken from~\cite{luterbacher1999reconstruction, luterbacher2001extending} and from December 1900 to July 2001 are derived from~\cite{trenberth1980northern}. Central European 500 year temperature reconstruction dataset, beginning from 1500 AD, is made available open source by NOAA National Centers for Environmental Information, under the World Data Service for Paleoclimatology. These were derived in the study~\cite{dobrovolny2010monthly}. We took winter only data points (months December, January and February) starting from the December of 1658 to the February of 2001 as it is known that the NAO influence is strongest in winter. This yielded a total of 1029 data points. However, reconstruction based on embedding was done for each year's winter separately (with a time delay of 1) and not in a continuous manner as for the other datasets, reducing the length of ordinal patterns encoded sequence to 343. Causal influence was checked between NAO and temperature for the encoded sequences using PCMI and PCCC and directly using one-dimensional CMI and CCC for the 1029 length sequences.

\emph{Daily NAO-temperature recordings:} Daily NAO records are available on the NOAA website and have been published in~\cite{barnston1987classification, chen2003sensitivity, van2000empirical}. Daily mean surface air temperature data from the Frankfurt station in Germany were taken from the records made available online by the ECA\&D project~\cite{klein2002daily}. This data was taken from 1\textsuperscript{st} January 1950 to 31\textsuperscript{st} April 2021. Once again, daily values from the winter months alone (December, January and February), comprising of 6390 data points, were extracted for the analysis. While embedding the two time series, care was taken not to embed the recordings of winter from one year along with that of winter from the next year. Causal influence was checked between daily winter NAO and temperature time-series.

\begin{figure}
\centering
\includegraphics[width=\columnwidth]{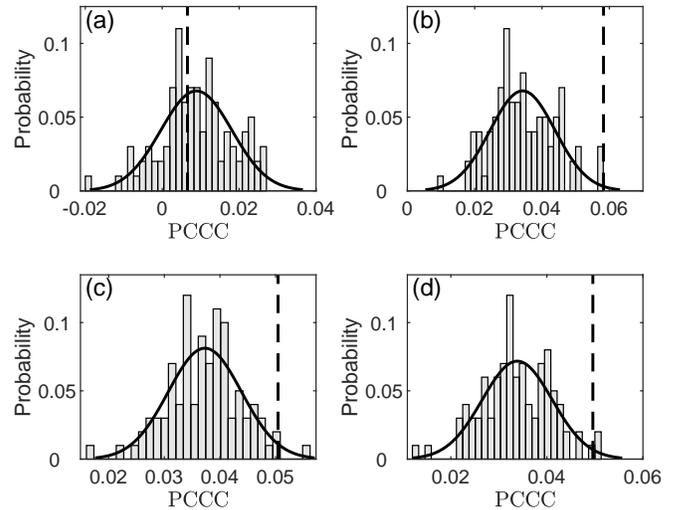}% Here is how to import EPS art
\caption{\label{fig_CCC_real_data} PCCC surrogate analysis results. PCCC surrogate analysis results for: (a) Kilo-year scale CO\textsubscript{2} $\rightarrow$ T, (b) Kilo-year scale T $\rightarrow$ CO\textsubscript{2}, (c) Yearly ENSO $\rightarrow$ SASM, (d) SASM $\rightarrow$ ENSO. Dashed line indicates PCCC value obtained for original series. Its position is indicated with respect to Gaussian curve fitted normalized histogram of surrogate PCCC values. PCCC for cases (b), (c), (d) is found to be significant.}
\end{figure}

For the analysis of causal interaction in each of these datasets, scalar CCC and CMI as well as PCCC and PCMI were computed as discussed in the `Methods' section. Parameters used for each of the methods are also given in the `Methods' section (Table~\ref{table_params}). In order to assess the significance of causality value estimated using each measure, 100 surrogate realizations were generated using the \emph{stationary bootstrap} method~\cite{politis1994stationary} for both the time series under consideration. Resampling of blocks of observations
of random length from the original time series is done for obtaining surrogate time series using this method. The length of each block has a geometric distribution. The probability parameter that determines the
geometric probability distribution for length of each block was set to 0.1 (as suggested in~\cite{politis1994stationary}). Significance testing of the causal interaction between original time-series was then done using a standard one-sided z-test, with p-value set to 0.05. Table~\ref{table_sig_caus} shows whether causal influence between the considered variables was found to be significant using each of the causality measures. Fig.~\ref{fig_CCC_real_data} depicts the value of the PCCC between original pair of time series with respect to the distribution of PCCC obtained using surrogate time series for two datasets: kilo-year scale CO\textsubscript{2}-temperature (Figs.~\ref{fig_CCC_real_data}(a) and~\ref{fig_CCC_real_data}(b)) and yearly scale ENSO-SASM (Figs.~\ref{fig_CCC_real_data}(c) and~\ref{fig_CCC_real_data}(d)) recordings. In the tables, Fig.~\ref{fig_CCC_real_data} and in the following text, we use the notation `T' to refer to temperature generically. Which of the temperature recordings is being referred to, will be clear from context.

\begin{table*}
\newcommand{\xmark}{\ding{55}}
\centering
\renewcommand{\arraystretch}{1.7}
\newcolumntype{C}[1]{>{\centering\arraybackslash}m{#1}}
\caption{Causal inference obtained for real datasets using different causality measures. \checkmark  indicates significant causality and \xmark  indicates non-significant causality.}
\label{table_sig_caus}
%\hspace{0.1\textwidth}
\begin{tabular}
{|C{5cm}|C{3cm}|C{1.5cm}|C{1.5cm}|C{1.5cm}|C{1.5cm}|}\specialrule{.15em}{.1em}{.1em}
%\backslashbox
\diagbox{{\bf System}}{{\bf Measure}} & {\bf Direction} & {\bf CCC} & {\bf PCCC} & {\bf CMI} & {\bf PCMI}\\ \specialrule{.15em}{.1em}{.1em}
\multirow{2}*{\textit{Millenial scale CO\textsubscript{2}-T}} & CO\textsubscript{2} $\rightarrow$ T & \checkmark & \xmark & \xmark & \xmark \\ \cline{2-6}
& T $\rightarrow$ CO\textsubscript{2} & \checkmark & \checkmark & \xmark & \xmark \\ \specialrule{.15em}{.1em}{.1em}
\multirow{2}*{\textit{Kilo-year scale CO\textsubscript{2}-T}} & CO\textsubscript{2} $\rightarrow$ T & \xmark & \xmark & \xmark & \xmark \\ \cline{2-6}
& T $\rightarrow$ CO\textsubscript{2} & \xmark & \checkmark & \xmark & \xmark \\ \specialrule{.15em}{.1em}{.1em}
\multirow{2}*{\textit{Kilo-year scale CH\textsubscript{4}-T}} & CH\textsubscript{4} $\rightarrow$ T & \xmark & \checkmark & \xmark & \xmark \\ \cline{2-6}
& T $\rightarrow$ CH\textsubscript{4} & \xmark & \xmark & \xmark & \xmark \\ \specialrule{.15em}{.1em}{.1em}
\multirow{2}*{\textit{Monthly scale CO\textsubscript{2}-T}} & CO\textsubscript{2} $\rightarrow$ T & \xmark & \checkmark & \xmark & \xmark \\ \cline{2-6}
& T $\rightarrow$ CO\textsubscript{2} & \xmark & \xmark & \xmark & \xmark \\ \specialrule{.15em}{.1em}{.1em}
\multirow{2}*{\textit{Yearly ENSO-SASM}} & ENSO $\rightarrow$ SASM & \xmark & \checkmark & \xmark & \xmark \\ \cline{2-6}
& SASM $\rightarrow$ ENSO & \checkmark & \checkmark & \xmark & \xmark \\ \specialrule{.15em}{.1em}{.1em}
\multirow{2}*{\textit{Monthly NINO-Indian monsoon}} & NINO $\rightarrow$ Monsoon & \xmark & \checkmark & \checkmark & \checkmark \\ \cline{2-6}
& Monsoon $\rightarrow$ NINO & \checkmark & \xmark & \checkmark & \checkmark \\ \specialrule{.15em}{.1em}{.1em}
\multirow{2}*{\textit{Monthly NAO-European T}} & NAO $\rightarrow$ T & \checkmark & \checkmark & \xmark & \xmark \\ \cline{2-6}
& T $\rightarrow$ NAO & \xmark & \xmark & \xmark & \xmark \\ \specialrule{.15em}{.1em}{.1em}
\multirow{2}*{\textit{Daily NAO-Frankfurt T}} & NAO $\rightarrow$ T & \checkmark & \xmark & \checkmark & \xmark \\ \cline{2-6}
& T $\rightarrow$ NAO & \xmark & \xmark & \xmark & \xmark \\ \specialrule{.15em}{.1em}{.1em}
\end{tabular}
\end{table*}

\section{Discussion and Conclusions}

CCC has been proposed as an `interventional' causality measure for time series. It does not require cause-effect separability in time series samples and is based on dynamical evolution of processes, making it suitable for subsampled time series, time series in which cause and effect are acquired at slightly different spatio-temporal scales than the scales at which they naturally occur and even when there are slight discrepancies in spatio-temporal scales of the cause and effect time series. This results in its robust performance in the case of missing samples, non-uniformly sampled, decimated and short length data~\cite{kathpalia2019data}. In this work, we have proposed the use of CCC in combination with ordinal pattern encoding. The latter preserves the dynamics of the systems of observed variables, allowing for CCC to decipher causal relationships between variables of multi-dimensional systems while conditioning for the presence other variables in these systems which might be unknown or unobserved.

Simulations of coupled R{\"o}ssler systems illustrate how scalar CCC is a complete failure for observables of coupled multi-dimensional dynamical systems, while PCCC performs well to determine the correct direction of coupling. Comparison of PCCC with PCMI for these simulations shows that the former beats the latter by showing better performance on shorter lengths of time series. Further, while PCMI consistently gave superior performance for increasing noise in coupled R{\"o}ssler systems, experiments with sparse data showed that PCCC outperforms PCMI. This was the case when samples were missing from the driver and response time series either in a synchronous or asynchronous manner.

As PCCC showed promising results for simulations with high levels of missing samples and short length, we have applied it to make causal inferences in datasets from climatology and paleoclimatology which suffer from the issues of irregular sampling, missing samples and (or) have limited number of data points available. Many of these datasets have been analyzed in previous studies. However, different studies report different results probably due to the challenging nature of their recordings available or the limitation of the inference methods applied to work on the data.

For example, the relationship between CO\textsubscript{2} concentrations and temperature of the atmosphere has been studied from the mid 1800s~\cite{foote1856art, arrhenius1896xxxi}, beginning when a strong link between the two was recognized. Relatively recently, with causal inference tools available, a number of studies have begun to look at the directionality of relationship between the two on different temporal scales. To mention a few findings, Kodra et al.~\cite{kodra2011exploring} found that CO\textsubscript{2} Granger causes temperature. Their analysis was based on data taken from 1860 to 2008. Atanassio~\cite{attanasio2012testing} found a clear evidence of GC from CO\textsubscript{2} to temperature using lag-augmented Wald test, for a similar time range. On the other hand, Stern and Kaufmann~\cite{stern2014anthropogenic} found bidirectional GC between the two, again for a similar time range. Kang and Larsson~\cite{kang2014link} also find bidirectional causation between the two using GC, however, by using data from ice cores for the last 800,000 years. Many of these latter studies criticize the former. Also, the drawbacks of one or more of these studies are explicitly mentioned in~\cite{triacca2001use, triacca2005granger, stern2014anthropogenic} and highlight the issues with the data and/ or the methodology employed. Other than GC and its extensions, a couple of other measures have also been used to study CO\textsubscript{2}-T relationship. Stips et al.~\cite{stips2016causal} have applied a measure called Liang's Information flow on CO\textsubscript{2}-T recordings, both on recent (1850-2005) and paleoclimate (800 ka ice core reconstructions) time-scales. The study finds unidirectional causation from CO\textsubscript{2} $\rightarrow$ T on the recent time-scale and from T $\rightarrow$ CO\textsubscript{2} on the paleoclimatic scale. They have also analysed the CH\textsubscript{4}-T relationship and found T to drive CH\textsubscript{4} on the paleoclimate scale. This study has been criticized by Goulet et al.~\cite{goulet2021spurious}. They show that an assumption of `linearity' made by Liang's information flow is nearly always rejected by the data. Convergent cross mapping, which is applied to the 800 ka recordings in another study, finds a bidirectional causal influence between both CO\textsubscript{2} - T and CH\textsubscript{4}-T~\cite{van2015causal}. Another recent study, that infers causation using lagged cross-correlations between monthly CO\textsubscript{2} and temperature, taken from the period 1980-2019, has found a bidirectional relationship on the recent monthly scale, with the dominant influence being from T $\rightarrow$ CO\textsubscript{2}~\cite{koutsoyiannis2020atmospheric}. In the light of the limitations of CCM~\cite{monster2017causal, schiecke2015convergent}, especially for irregularly sampled or missing data~\cite{kathpalia2021theoretical}, and of the widely known pitfalls of correlation coefficient~\cite{janse2021conducting}, it is difficult to rely on the inferences of the latter two studies.

PCCC indicates unidirectional causality from T $\rightarrow$ CO\textsubscript{2} on the paleoclimatic scale, using both millenial and kilo-year scale recordings. On the recent monthly scale, the situation is reversed with CO\textsubscript{2} driving T. These results are in line with some of the existing CO\textsubscript{2}-T causal analysis studies and clearly PCCC does not suffer the limitations of existing approaches. On the kilo-year scale, PCCC suggests that CH\textsubscript{4} drives T. While none of the above discussed causality studies have found this result, other works have suggested that methane concentrations modulate millenial-scale climate variability because of the sensitivity of methane to insolation~\cite{brook1996rapid, thirumalai2020methane}. Other approaches implemented in this study -- CCC, CMI, PCMI also do not duplicate the results obtained by PCCC because of their specific limitations such as the inability to work on multi-dimensional, short length or irregularly sampled data.

ENSO events and the Indian monsoon are other major climatic processes of global importance~\cite{solomon2007climate}. The relationship between the two has been studied extensively, especially using correlation and coherence approaches~\cite{kripalani1997rainfall, kumar1999weakening, krishnamurthy2000indian, sarkar2004further, maraun2005epochs, zubair2006strengthening}. While ENSO is normally expected to play a driving role, there is no clear consensus on the directionality of the relationship between the two processes. More recently, causal inference approaches have been used to study the nature of their coupling. In~\cite{mokhov2011alternating} and \cite{mokhov2012relationship}, both linear and non-linear GC versions were implemented on monthly mean ENSO-Indian monsoon time series, ranging from the period 1871-2006 and bidirectional coupling was inferred between the two processes. Other studies have studied the causal relationship indirectly by analyzing the ENSO-Indian Ocean Dipole link. For example, in~\cite{le2020causal}, this connection was studied by applying GC on yearly reanalysis as well as model data ranging from 1950-2014. The study found robust causal influence of Indian Ocean Dipole on ENSO while the influence in opposite direction had lower confidence. Using PCCC, we find a bidirectional causal influence between yearly recordings of ENSO-SASM. However, on the shorter monthly scales, NINO is found to drive Indian Monsoon and there is insignificant effect in the opposite direction.

Although the NAO is known to be a leading mode of winter climate variability over Europe~\cite{wanner2001north, hurrell2010north, deser2017role}, the directionality or feedback in NAO related climate effects has been studied by a few causality analysis studies~\cite{wang2004relation, mosedale2006granger, wang2019central}. We investigate the NAO-European temperatures relationship on both monthly and daily time scales using winter only data. While PCCC indicates that NAO drives central European temperatures with no significant feedback on the longer monthly scale, on the daily scale it shows no significant causation in either direction. On the other hand, CCC and CMI, based on one dimensional time series, indicate a strong influence from NAO to Frankfurt daily mean temperatures. This result indicates that the NAO influence on European winter temperature on the daily scale can be explained as a simple time-delayed transfer of information between scalar time series in which no role is played by higher-dimensional patterns, potentially reflected in ordinal coding. Such an information transfer in the atmosphere is tied to the transfer of mass and energy as indicated in the study of climate networks by Hlinka et al.~\cite{hlinka2017smooth}. CMI and PCMI estimates can be considered to be reliable for this analysis as the time-series analyzed are long, close to 6000 time points.

CCC is free of the assumptions of linearity, requirement of long-term stationarity, extremely robust to missing samples, irregular sampling and short length data; and its combination with permutation patterns allows it to make reliable inferences for coupled systems with multiple variables. Thus, we can expect our analysis and inferences presented here on some highly-researched and long-debated climatic interactions to be highly robust and reliable. We also expect that the use of PCCC on other challenging datasets from climatology and other fields will be helpful to shed light on the causal linkages in considered systems.

\section{Methods}

{\bf Compression Complexity Causality (CCC)} is defined as the change in the dynamical compression-complexity of time series $y$ when $\Delta y$ is seen to be generated jointly by the dynamical evolution of both $y_{past}$ and $x_{past}$ as opposed to by the reality of the dynamical evolution of $y_{past}$ alone. $y_{past}, x_{past}$ are windows of a particular length $L$ taken from contemporary time points of time series $y$ and $x$ respectively and $\Delta y$ is a window of length $w$ following $y_{past}$~\cite{kathpalia2019data}. Dynamical compression-complexity (CC) is estimated using the measure effort-to-compress (ETC)~\cite{nagaraj2013new} and given by:

\begin{equation}
\label{eq_complexity_1D}
CC(\Delta y \vert y_{past} )=ETC(y_{past}+\Delta y)-ETC(y_{past} ),
\end{equation}

\begin{equation}
\begin{split}
& CC(\Delta y \vert y_{past}, x_{past} ) = \\ & ETC(y_{past}+\Delta y, x_{past}+\Delta y) - ETC(x_{past},y_{past} ),
\label{eq_complexity_2D}
\end{split}
\end{equation}

Eq.~(\ref{eq_complexity_1D}) computes the dynamical compression-complexity of $\Delta y$ as a dynamical evolution of $y_{past}$ alone. Eq.~(\ref{eq_complexity_2D}) computes the dynamical compression-complexity of $\Delta y$ as a dynamical evolution of both $y_{past}$ and $x_{past}$. $CCC_{x_{past} \rightarrow y}$ is then estimated as:

\begin{equation}
\label{eq_ind_CCC}
CCC_{x_{past} \rightarrow \Delta y}= CC(\Delta y \vert y_{past} ) - CC(\Delta y \vert y_{past}, x_{past} ).
\end{equation}

Averaged CCC from $x$ to $y$ over the entire length of time series with the window $\Delta y$ being slided by a step-size of $\delta$ is estimated as ---
\begin{equation}
\begin{split}
\label{eq_CCC}
CCC_{x \rightarrow y} & =  \overline{CCC}_{x_{past} \rightarrow \Delta y} \\ &= \overline{CC}(\Delta y \vert y_{past} ) - \overline{CC}(\Delta y \vert x_{past}, y_{past} ),
\end{split}
\end{equation}

If $\overline{CC}(\Delta y \vert y_{past} ) \approx \overline{CC}(\Delta y \vert x_{past}, y_{past})$, there is no causality from $x$ to $y$. Surrogate time series are generated for both $x$ and $y$ and the $CCC_{x \rightarrow y}$ values of the original and surrogate time series compared. If the CCC computed for original time series is statistically different from that of surrogate time series, we can infer the presence of causal relation from $x \rightarrow y$~\cite{kathpalia2021theoretical}. $CCC_{x \rightarrow y}$ can be both $<$ or $> 0$ depending upon the \emph{nature} or \emph{quality} of the causal relationship~\cite{kathpalia2019data}. The magnitude indicates the strength of causation.

Selection of parameters: $L,w,\delta$ and the number of bins, $B$, for symbolizing the time series using equidistant binning (ETC is applied to symbolic sequences) is done using parameter selection criteria given in the supplementary text of~\cite{kathpalia2019data}.

{\bf Permutation Compression-Complexity Causality} is the causal inference technique proposed and implemented in this work. Given a pair of time series $x_1$ and $x_2$ from dynamical systems in which causation is to be checked from $x_1$ to $x_2$, we first embed the time series of the potential driver ($x_1$ here) in the following manner: $x_1(t), x_1(t+\eta), x_1(t+2\eta), \ldots x_1(t+(m-1)\eta)$, where $\eta$ is the time delay and $m$ is the embedding dimension of $x_1$. $\eta$ is computed as the first minimum of auto mutual information function. The embedded time-series at each time-point is then symbolized using permutation or ordinal patterns binning. For example, if $m=3$, the embedding at time point $t$ is given as $\hat{x}_1(t)=(x_1(t), x_1(t+\eta), x_1(t+2\eta))$. Symbols $0,1,2$ are then used for labelling the pattern for $\hat{x}(t)$ at each time point by sorting the embedded values in ascending order, with 2 being used for the highest value and 0 for the lowest. If two or more values are exactly same, they are labelled differently depending on the order of their occurrence. A total of $m!=3!$ patterns at time $t$ are possible in this case. Thus, $\hat{x}(t)$ is symbolized to a one dimensional sequence consisting of $m!$ possible symbols or bins. $CCC$ is then estimated from $\hat{x}_1(t)$ to $x_2(t)$, using Eq.~(\ref{eq_CCC}) after symbolizing $x_2(t)$ using standard equidistant binning with $m!$ bins. Thus,
\begin{equation}
    PCCC_{x_1 \rightarrow x_2}=CCC_{\hat{x}_1 \rightarrow x_2}.
    \label{eq_PCCC}
\end{equation}

Permutation binning is not done for the potential driver series as it was found from simulation experiments (R{\"o}ssler data) that embedding the `cause' alone works better for the CCC measure. Full dimensionality of the cause is necessary to predict the effect. Hence, embedding only the cause helps to recover the causal relationship. PCCC helps to take into account the multidimensional nature of the coupled systems. Parameter selection for PCCC is done in the same manner as for the case of CCC, using the symbolic sequences, $\hat{x}_1(t)$ and $x_2(t)$, for selection of the parameters. When PCCC is to be estimated from $x_2 \rightarrow x_1$, $x_2$ is embedded and $x_1$ remains as it is. Just like CCC, the PCCC measure can also take negative values.

{\bf Conditional Mutual Information (CMI)} of the variables $X$ and $Y$ given the variable $Z$ is a common information-theoretic functional used for the causality detection, and can be obtained as

\begin{equation}
I(X;Y|Z)=H(X|Z)+H(Y|Z)-H(X,Y|Z)
    \label{eq_CMI}
\end{equation}

where $H(X_1,X_2,...|Z)=H(X_1,X_2,...)-H(Z)$ is the conditional entropy, and  the joint Shannon entropy $H(X_1,X_2,...)$ is defined as:

\begin{equation}
H(X_1,X_2,...)=-\sum_{x_1,x_2,...}{p(x_1,x_2,...)\log{p(x_1,x_2,...)}}
\label{H}
\end{equation}
where $p(x_1,x_2,...)=Pr[X_1=x_1,X_2=x_2,...]$ is the joint probability distribution function of the amplitude of variables $\{X_1,X_2,...\}$. In order to detect the coupling direction among two dynamical variables of $X$ and $Y$, Palu\v{s} et al. \cite{paluvs2001synchronization} used the conditional mutual information $I(X(t);Y(t+\tau)|Y(t))$, that captures the net information about the $\tau$-future of the process $Y$ contained in the process $X$. As mentioned in the Introduction, to estimate other unknown variables, an $m$-dimensional state vector $X$ can be reconstructed as $X(t)=\{x(t),x(t-\eta),...,x(t-(m-1)\eta)\}$. Accordingly, CMI defined above can be represented by its reconstructed version for all variables of $X(t)$, $Y(t+\tau)$ and $Y(t)$. However, extensive numerical studies \cite{paluvs2007directionality} demonstrated that CMI in the form

\begin{equation}
I(X(t);Y(t+\tau)|Y(t),Y(t-\eta),...,Y(t-(m-1)\eta))
\end{equation}

is sufficient to infer direction of coupling among dynamical variables of $X(t)$ and $Y(t)$. In this respect, we use this measure to detect causality relationships in this article.

{\bf Permutation Conditional Mutual Information (PCMI)} can be obtained based on the permutation analysis described earlier in the PCCC definition. In this approach, all marginal, joint or conditional probability distribution functions of the amplitude of the variables are replaced by their symbolized versions, thus Eq.~(\ref{H}) should be replaced by
\begin{equation}
H(\hat{X}_1,\hat{X}_2,...)=-\sum_{\hat{x}_1,\hat{x}_2,...}{p(\hat{x}_1,\hat{x}_2,...)\log{p(\hat{x}_1,\hat{x}_2,...)}}
\label{symb_H}
\end{equation}
where $p(\hat{x}_1,\hat{x}_2,...)=Pr[\hat{X}_1=\hat{x}_1,\hat{X}_2=\hat{x}_2,...]$ is the joint probability distribution function of the symbolized variables $\hat{X}_i(t)=\{X_i(t),X_i(t+\eta),...,X_i(t+(m-1)\eta)\}$. By using Eqs.~(\ref{eq_CMI}) and (\ref{symb_H}), permutation CMI can be obtained as $I(\hat{X}(t);\hat{Y}(t+\tau)|\hat{Y}(t))$. Finally, one should replace $\tau$ with $\tau+(m-1)\eta$ in order to avoid any overlapping between the past and future of the symbolized variable $\hat{Y}$.

%{\bf Parameters used for different real-datasets}
{\bf Parameters of the methods} used were set as shown in Table~\ref{table_params} for different datasets.

\begin{table}
\caption{Parameters corresponding to each method, used for different datasets.}
\label{table_params}
\centering
\renewcommand{\arraystretch}{2}
\newcolumntype{C}[1]{>{\centering\arraybackslash}m{#1}}
\begin{tabular}{|C{1.7cm}|C{1.8cm}|C{1.3cm}|C{1.3cm}|C{1.5cm}|}\hline
{\bf Dataset} & {\bf Embedding} & {\bf CCC} & {\bf PCCC} & {\bf CMI}/ {\bf PCMI} \\ \hline
{\it R{\"o}ssler} & $\eta_{x_1}=5$ $\eta_{x_2}=5$ $m=3$ & $L=300$ $w=30$ $\delta=30$ $B=8$ & $L=25$ $w=15$ $\delta=20$ & $\tau=20$ \\ \hline
{\it Millenial CO\textsubscript{2}-T} & $\eta_{CO_2}=11$ $\eta_{T}=16$ $m=3$ & $L=60$ $w=15$ $\delta=20$ $B=4$ & $L=60$ $w=30$ $\delta=20$ & $\tau=1-30$ \\ \hline
{\it Kilo-year CO\textsubscript{2}-T} & $\eta_{CO_2}=24$ $\eta_{T}=8$ $m=3$ & $L=60$ $w=15$ $\delta=20$ $B=4$ & $L=30$ $w=15$ $\delta=20$ & $\tau=1-30$ \\ \hline
{\it Kilo-year CH\textsubscript{4}-T} & $\eta_{CH_4}=10$ $\eta_{T}=8$ $m=3$ & $L=60$ $w=15$ $\delta=20$ $B=4$ & $L=30$ $w=15$ $\delta=20$ & $\tau=1-30$ \\ \hline
{\it Monthly CO\textsubscript{2}-T} & $\eta_{CO_2}=3$ $\eta_{T}=2$ $m=3$ & $L=60$ $w=15$ $\delta=20$ $B=4$ & $L=30$ $w=15$ $\delta=20$ & $\tau=1-30$ \\ \hline
{\it Yearly ENSO-SASM} & $\eta_{ENSO}=1$ $\eta_{SASM}=4$ $m=3$ & $L=60$ $w=15$ $\delta=20$ $B=4$ & $L=60$ $w=30$ $\delta=30$ & $\tau=1-30$ \\ \hline
{\it Monthly NINO-India Monsoon} & $\eta_{NINO}=10$ $\eta_{mon}=3$ $m=3$ & $L=60$ $w=15$ $\delta=20$ $B=4$ & $L=30$ $w=15$ $\delta=20$ & $\tau=1-30$ \\ \hline
{\it Monthly NAO-T} & $\eta_{NAO}=1$ $\eta_{T}=1$ $m=3$ & $L=60$ $w=15$ $\delta=20$ $B=4$ & $L=30$ $w=15$ $\delta=10$ & $\tau=1-30$ \\ \hline
{\it Daily NAO-T} & $\eta_{NAO}=15$ $\eta_{T}=15$ $m=3$ & $L=40$ $w=15$ $\delta=20$ $B=4$ & $L=30$ $w=15$ $\delta=20$ & $\tau=1-30$ \\ \hline
\end{tabular}
\end{table}

\section*{Data Availability}
The millenial scale CO\textsubscript{2} and temperature datasets are freely available at \url{https://zenodo.org/record/4562996#.YiDbTN_ML3A}. Kilo-year scale CO\textsubscript{2}, CH\textsubscript{4} and temperature datasets are available as supplementary files for~\cite{past2016interglacials} at \url{https://agupubs.onlinelibrary.wiley.com/doi/full/10.1002/2015RG000482}. Monthly CO\textsubscript{2} recordings are taken from the NOAA repository and are available at \url{https://gml.noaa.gov/ccgg/trends/}. Monthly Northern hemisphere temperature anomaly recordings are taken from the NOAA repository and are available at \url{https://www.ncdc.noaa.gov/cag/global/time-series}. The yearly El Niño/Southern Oscillation Index Reconstruction dataset is taken from the NOAA repository, \url{https://www.ncei.noaa.gov/access/paleo-search/study/11194}. The yearly South Asian Summer Monsoon Index Reconstruction dataset is taken from the NOAA repository, \url{https://www.ncei.noaa.gov/access/paleo-search/study/17369}. Monthly Niño 3.4 SST Index dataset is taken from the NOAA repository, available at \url{https://psl.noaa.gov/gcos_wgsp/Timeseries/Nino34/}. Monthly all India rainfall dataset is made available by the World Metereological Organization at \url{http://climexp.knmi.nl/data/pALLIN.dat}. Reconstructed
monthly North Atlantic Oscillation Index is available at the NOAA repository, \url{https://psl.noaa.gov/gcos_wgsp/Timeseries/RNAO/}. Monthly Central European 500 Year Temperature Reconstructions are available at the NOAA repository, \url{https://www.ncei.noaa.gov/access/metadata/landing-page/bin/iso?id=noaa-recon-9970}. Daily North Atlantic Oscillation Index is available at the NOAA repository, \url{https://www.cpc.ncep.noaa.gov/products/precip/CWlink/pna/nao.shtml}. Daily Frankfurt air temperatures are made available by the ECA\&D project at \url{https://www.ecad.eu/dailydata/predefinedseries.php}.

\section*{Code Availability}
The computer codes used in this study are freely available at \url{https://github.com/AditiKathpalia/PermutationCCC}
under the Apache 2.0 Open-source license.

\begin{acknowledgments}
This study is supported by the Czech Science Foundation, Project No.~GA19-16066S and by the Czech Academy of Sciences, Praemium Academiae awarded to M. Palu\v{s}.
\end{acknowledgments}

%\nocite{*}

\bibliography{apssamp}% Produces the bibliography via BibTeX.

\providecommand{\noopsort}[1]{}\providecommand{\singleletter}[1]{#1}%
\begin{thebibliography}{115}%
\makeatletter
\providecommand \@ifxundefined [1]{%
 \@ifx{#1\undefined}
}%
\providecommand \@ifnum [1]{%
 \ifnum #1\expandafter \@firstoftwo
 \else \expandafter \@secondoftwo
 \fi
}%
\providecommand \@ifx [1]{%
 \ifx #1\expandafter \@firstoftwo
 \else \expandafter \@secondoftwo
 \fi
}%
\providecommand \natexlab [1]{#1}%
\providecommand \enquote  [1]{``#1''}%
\providecommand \bibnamefont  [1]{#1}%
\providecommand \bibfnamefont [1]{#1}%
\providecommand \citenamefont [1]{#1}%
\providecommand \href@noop [0]{\@secondoftwo}%
\providecommand \href [0]{\begingroup \@sanitize@url \@href}%
\providecommand \@href[1]{\@@startlink{#1}\@@href}%
\providecommand \@@href[1]{\endgroup#1\@@endlink}%
\providecommand \@sanitize@url [0]{\catcode `\\12\catcode `\$12\catcode
  `\&12\catcode `\#12\catcode `\^12\catcode `\_12\catcode `\%12\relax}%
\providecommand \@@startlink[1]{}%
\providecommand \@@endlink[0]{}%
\providecommand \url  [0]{\begingroup\@sanitize@url \@url }%
\providecommand \@url [1]{\endgroup\@href {#1}{\urlprefix }}%
\providecommand \urlprefix  [0]{URL }%
\providecommand \Eprint [0]{\href }%
\providecommand \doibase [0]{https://doi.org/}%
\providecommand \selectlanguage [0]{\@gobble}%
\providecommand \bibinfo  [0]{\@secondoftwo}%
\providecommand \bibfield  [0]{\@secondoftwo}%
\providecommand \translation [1]{[#1]}%
\providecommand \BibitemOpen [0]{}%
\providecommand \bibitemStop [0]{}%
\providecommand \bibitemNoStop [0]{.\EOS\space}%
\providecommand \EOS [0]{\spacefactor3000\relax}%
\providecommand \BibitemShut  [1]{\csname bibitem#1\endcsname}%
\let\auto@bib@innerbib\@empty
%</preamble>
\bibitem [{\citenamefont {Pearl}\ and\ \citenamefont
  {Mackenzie}(2018)}]{pearl2018book}%
  \BibitemOpen
  \bibfield  {author} {\bibinfo {author} {\bibfnamefont {J.}~\bibnamefont
  {Pearl}}\ and\ \bibinfo {author} {\bibfnamefont {D.}~\bibnamefont
  {Mackenzie}},\ }\href@noop {} {\emph {\bibinfo {title} {The book of why: the
  new science of cause and effect}}}\ (\bibinfo  {publisher} {Basic Books},\
  \bibinfo {year} {2018})\BibitemShut {NoStop}%
\bibitem [{\citenamefont {Kathpalia}\ and\ \citenamefont
  {Nagaraj}(2021)}]{kathpalia2021measuring}%
  \BibitemOpen
  \bibfield  {author} {\bibinfo {author} {\bibfnamefont {A.}~\bibnamefont
  {Kathpalia}}\ and\ \bibinfo {author} {\bibfnamefont {N.}~\bibnamefont
  {Nagaraj}},\ }\bibfield  {title} {\bibinfo {title} {Measuring causality},\
  }\href@noop {} {\bibfield  {journal} {\bibinfo  {journal} {Resonance}\ ,\
  \bibinfo {pages} {191}} (\bibinfo {year} {2021})}\BibitemShut {NoStop}%
\bibitem [{\citenamefont {Wiener}(1956)}]{wiener}%
  \BibitemOpen
  \bibfield  {author} {\bibinfo {author} {\bibfnamefont {N.}~\bibnamefont
  {Wiener}},\ }\bibfield  {title} {\bibinfo {title} {The theory of
  prediction},\ }\href@noop {} {\bibfield  {journal} {\bibinfo  {journal}
  {Modern mathematics for engineers}\ }\textbf {\bibinfo {volume} {1}},\
  \bibinfo {pages} {125} (\bibinfo {year} {1956})}\BibitemShut {NoStop}%
\bibitem [{\citenamefont {Granger}(1969)}]{granger}%
  \BibitemOpen
  \bibfield  {author} {\bibinfo {author} {\bibfnamefont {C.}~\bibnamefont
  {Granger}},\ }\bibfield  {title} {\bibinfo {title} {Investigating causal
  relations by econometric models and cross-spectral methods},\ }\href@noop {}
  {\bibfield  {journal} {\bibinfo  {journal} {Econometrica}\ }\textbf {\bibinfo
  {volume} {37}},\ \bibinfo {pages} {424–} (\bibinfo {year}
  {1969})}\BibitemShut {NoStop}%
\bibitem [{\citenamefont {Geweke}(1984)}]{geweke1984inference}%
  \BibitemOpen
  \bibfield  {author} {\bibinfo {author} {\bibfnamefont {J.}~\bibnamefont
  {Geweke}},\ }\bibfield  {title} {\bibinfo {title} {Inference and causality in
  economic time series models},\ }\href@noop {} {\bibfield  {journal} {\bibinfo
   {journal} {Handbook of econometrics}\ }\textbf {\bibinfo {volume} {2}},\
  \bibinfo {pages} {1101} (\bibinfo {year} {1984})}\BibitemShut {NoStop}%
\bibitem [{\citenamefont {Hiemstra}\ and\ \citenamefont
  {Jones}(1994)}]{hiemstra1994testing}%
  \BibitemOpen
  \bibfield  {author} {\bibinfo {author} {\bibfnamefont {C.}~\bibnamefont
  {Hiemstra}}\ and\ \bibinfo {author} {\bibfnamefont {J.~D.}\ \bibnamefont
  {Jones}},\ }\bibfield  {title} {\bibinfo {title} {Testing for linear and
  nonlinear granger causality in the stock price-volume relation},\ }\href@noop
  {} {\bibfield  {journal} {\bibinfo  {journal} {The Journal of Finance}\
  }\textbf {\bibinfo {volume} {49}},\ \bibinfo {pages} {1639} (\bibinfo {year}
  {1994})}\BibitemShut {NoStop}%
\bibitem [{\citenamefont {Chiou-Wei}\ \emph {et~al.}(2008)\citenamefont
  {Chiou-Wei}, \citenamefont {Chen},\ and\ \citenamefont
  {Zhu}}]{chiou2008economic}%
  \BibitemOpen
  \bibfield  {author} {\bibinfo {author} {\bibfnamefont {S.~Z.}\ \bibnamefont
  {Chiou-Wei}}, \bibinfo {author} {\bibfnamefont {C.-F.}\ \bibnamefont
  {Chen}},\ and\ \bibinfo {author} {\bibfnamefont {Z.}~\bibnamefont {Zhu}},\
  }\bibfield  {title} {\bibinfo {title} {Economic growth and energy consumption
  revisited: evidence from linear and nonlinear granger causality},\
  }\href@noop {} {\bibfield  {journal} {\bibinfo  {journal} {Energy Economics}\
  }\textbf {\bibinfo {volume} {30}},\ \bibinfo {pages} {3063} (\bibinfo {year}
  {2008})}\BibitemShut {NoStop}%
\bibitem [{\citenamefont {Seth}\ \emph {et~al.}(2015)\citenamefont {Seth},
  \citenamefont {Barrett},\ and\ \citenamefont {Barnett}}]{seth2015granger}%
  \BibitemOpen
  \bibfield  {author} {\bibinfo {author} {\bibfnamefont {A.~K.}\ \bibnamefont
  {Seth}}, \bibinfo {author} {\bibfnamefont {A.~B.}\ \bibnamefont {Barrett}},\
  and\ \bibinfo {author} {\bibfnamefont {L.}~\bibnamefont {Barnett}},\
  }\bibfield  {title} {\bibinfo {title} {Granger causality analysis in
  neuroscience and neuroimaging},\ }\href@noop {} {\bibfield  {journal}
  {\bibinfo  {journal} {Journal of Neuroscience}\ }\textbf {\bibinfo {volume}
  {35}},\ \bibinfo {pages} {3293} (\bibinfo {year} {2015})}\BibitemShut
  {NoStop}%
\bibitem [{\citenamefont {Mosedale}\ \emph {et~al.}(2006)\citenamefont
  {Mosedale}, \citenamefont {Stephenson}, \citenamefont {Collins},\ and\
  \citenamefont {Mills}}]{mosedale2006granger}%
  \BibitemOpen
  \bibfield  {author} {\bibinfo {author} {\bibfnamefont {T.~J.}\ \bibnamefont
  {Mosedale}}, \bibinfo {author} {\bibfnamefont {D.~B.}\ \bibnamefont
  {Stephenson}}, \bibinfo {author} {\bibfnamefont {M.}~\bibnamefont
  {Collins}},\ and\ \bibinfo {author} {\bibfnamefont {T.~C.}\ \bibnamefont
  {Mills}},\ }\bibfield  {title} {\bibinfo {title} {Granger causality of
  coupled climate processes: Ocean feedback on the north atlantic
  oscillation},\ }\href@noop {} {\bibfield  {journal} {\bibinfo  {journal}
  {Journal of climate}\ }\textbf {\bibinfo {volume} {19}},\ \bibinfo {pages}
  {1182} (\bibinfo {year} {2006})}\BibitemShut {NoStop}%
\bibitem [{\citenamefont {Tirabassi}\ \emph {et~al.}(2015)\citenamefont
  {Tirabassi}, \citenamefont {Masoller},\ and\ \citenamefont
  {Barreiro}}]{tirabassi2015study}%
  \BibitemOpen
  \bibfield  {author} {\bibinfo {author} {\bibfnamefont {G.}~\bibnamefont
  {Tirabassi}}, \bibinfo {author} {\bibfnamefont {C.}~\bibnamefont
  {Masoller}},\ and\ \bibinfo {author} {\bibfnamefont {M.}~\bibnamefont
  {Barreiro}},\ }\bibfield  {title} {\bibinfo {title} {A study of the air--sea
  interaction in the south atlantic convergence zone through granger
  causality},\ }\href@noop {} {\bibfield  {journal} {\bibinfo  {journal}
  {International Journal of Climatology}\ }\textbf {\bibinfo {volume} {35}},\
  \bibinfo {pages} {3440} (\bibinfo {year} {2015})}\BibitemShut {NoStop}%
\bibitem [{\citenamefont {Runge}\ \emph {et~al.}(2019)\citenamefont {Runge},
  \citenamefont {Bathiany}, \citenamefont {Bollt}, \citenamefont {Camps-Valls},
  \citenamefont {Coumou}, \citenamefont {Deyle}, \citenamefont {Glymour},
  \citenamefont {Kretschmer}, \citenamefont {Mahecha}, \citenamefont
  {Mu{\~n}oz-Mar{\'\i}} \emph {et~al.}}]{runge2019inferring}%
  \BibitemOpen
  \bibfield  {author} {\bibinfo {author} {\bibfnamefont {J.}~\bibnamefont
  {Runge}}, \bibinfo {author} {\bibfnamefont {S.}~\bibnamefont {Bathiany}},
  \bibinfo {author} {\bibfnamefont {E.}~\bibnamefont {Bollt}}, \bibinfo
  {author} {\bibfnamefont {G.}~\bibnamefont {Camps-Valls}}, \bibinfo {author}
  {\bibfnamefont {D.}~\bibnamefont {Coumou}}, \bibinfo {author} {\bibfnamefont
  {E.}~\bibnamefont {Deyle}}, \bibinfo {author} {\bibfnamefont
  {C.}~\bibnamefont {Glymour}}, \bibinfo {author} {\bibfnamefont
  {M.}~\bibnamefont {Kretschmer}}, \bibinfo {author} {\bibfnamefont {M.~D.}\
  \bibnamefont {Mahecha}}, \bibinfo {author} {\bibfnamefont {J.}~\bibnamefont
  {Mu{\~n}oz-Mar{\'\i}}}, \emph {et~al.},\ }\bibfield  {title} {\bibinfo
  {title} {Inferring causation from time series in earth system sciences},\
  }\href@noop {} {\bibfield  {journal} {\bibinfo  {journal} {Nature
  communications}\ }\textbf {\bibinfo {volume} {10}},\ \bibinfo {pages} {1}
  (\bibinfo {year} {2019})}\BibitemShut {NoStop}%
\bibitem [{\citenamefont {Bell}\ \emph {et~al.}(1996)\citenamefont {Bell},
  \citenamefont {Kay},\ and\ \citenamefont {Malley}}]{bell1996non}%
  \BibitemOpen
  \bibfield  {author} {\bibinfo {author} {\bibfnamefont {D.}~\bibnamefont
  {Bell}}, \bibinfo {author} {\bibfnamefont {J.}~\bibnamefont {Kay}},\ and\
  \bibinfo {author} {\bibfnamefont {J.}~\bibnamefont {Malley}},\ }\bibfield
  {title} {\bibinfo {title} {A non-parametric approach to non-linear causality
  testing},\ }\href@noop {} {\bibfield  {journal} {\bibinfo  {journal}
  {Economics Letters}\ }\textbf {\bibinfo {volume} {51}},\ \bibinfo {pages} {7}
  (\bibinfo {year} {1996})}\BibitemShut {NoStop}%
\bibitem [{\citenamefont {Chen}\ \emph {et~al.}(2004)\citenamefont {Chen},
  \citenamefont {Rangarajan}, \citenamefont {Feng},\ and\ \citenamefont
  {Ding}}]{chen2004analyzing}%
  \BibitemOpen
  \bibfield  {author} {\bibinfo {author} {\bibfnamefont {Y.}~\bibnamefont
  {Chen}}, \bibinfo {author} {\bibfnamefont {G.}~\bibnamefont {Rangarajan}},
  \bibinfo {author} {\bibfnamefont {J.}~\bibnamefont {Feng}},\ and\ \bibinfo
  {author} {\bibfnamefont {M.}~\bibnamefont {Ding}},\ }\bibfield  {title}
  {\bibinfo {title} {Analyzing multiple nonlinear time series with extended
  granger causality},\ }\href@noop {} {\bibfield  {journal} {\bibinfo
  {journal} {Physics letters A}\ }\textbf {\bibinfo {volume} {324}},\ \bibinfo
  {pages} {26} (\bibinfo {year} {2004})}\BibitemShut {NoStop}%
\bibitem [{\citenamefont {Schiff}\ \emph {et~al.}(1996)\citenamefont {Schiff},
  \citenamefont {So}, \citenamefont {Chang}, \citenamefont {Burke},\ and\
  \citenamefont {Sauer}}]{schiff1996detecting}%
  \BibitemOpen
  \bibfield  {author} {\bibinfo {author} {\bibfnamefont {S.~J.}\ \bibnamefont
  {Schiff}}, \bibinfo {author} {\bibfnamefont {P.}~\bibnamefont {So}}, \bibinfo
  {author} {\bibfnamefont {T.}~\bibnamefont {Chang}}, \bibinfo {author}
  {\bibfnamefont {R.~E.}\ \bibnamefont {Burke}},\ and\ \bibinfo {author}
  {\bibfnamefont {T.}~\bibnamefont {Sauer}},\ }\bibfield  {title} {\bibinfo
  {title} {Detecting dynamical interdependence and generalized synchrony
  through mutual prediction in a neural ensemble},\ }\href@noop {} {\bibfield
  {journal} {\bibinfo  {journal} {Physical Review E}\ }\textbf {\bibinfo
  {volume} {54}},\ \bibinfo {pages} {6708} (\bibinfo {year}
  {1996})}\BibitemShut {NoStop}%
\bibitem [{\citenamefont {Le~Van~Quyen}\ \emph {et~al.}(1999)\citenamefont
  {Le~Van~Quyen}, \citenamefont {Martinerie}, \citenamefont {Adam},\ and\
  \citenamefont {Varela}}]{le1999nonlinear}%
  \BibitemOpen
  \bibfield  {author} {\bibinfo {author} {\bibfnamefont {M.}~\bibnamefont
  {Le~Van~Quyen}}, \bibinfo {author} {\bibfnamefont {J.}~\bibnamefont
  {Martinerie}}, \bibinfo {author} {\bibfnamefont {C.}~\bibnamefont {Adam}},\
  and\ \bibinfo {author} {\bibfnamefont {F.~J.}\ \bibnamefont {Varela}},\
  }\bibfield  {title} {\bibinfo {title} {Nonlinear analyses of interictal eeg
  map the brain interdependences in human focal epilepsy},\ }\href@noop {}
  {\bibfield  {journal} {\bibinfo  {journal} {Physica D: Nonlinear Phenomena}\
  }\textbf {\bibinfo {volume} {127}},\ \bibinfo {pages} {250} (\bibinfo {year}
  {1999})}\BibitemShut {NoStop}%
\bibitem [{\citenamefont {Marinazzo}\ \emph {et~al.}(2008)\citenamefont
  {Marinazzo}, \citenamefont {Pellicoro},\ and\ \citenamefont
  {Stramaglia}}]{marinazzo}%
  \BibitemOpen
  \bibfield  {author} {\bibinfo {author} {\bibfnamefont {D.}~\bibnamefont
  {Marinazzo}}, \bibinfo {author} {\bibfnamefont {M.}~\bibnamefont
  {Pellicoro}},\ and\ \bibinfo {author} {\bibfnamefont {S.}~\bibnamefont
  {Stramaglia}},\ }\bibfield  {title} {\bibinfo {title} {Kernel method for
  nonlinear granger causality},\ }\href@noop {} {\bibfield  {journal} {\bibinfo
   {journal} {Physical Review Letters}\ }\textbf {\bibinfo {volume} {100}},\
  \bibinfo {pages} {144103} (\bibinfo {year} {2008})}\BibitemShut {NoStop}%
\bibitem [{\citenamefont {Baccal{\'a}}\ and\ \citenamefont
  {Sameshima}(2001)}]{baccala2001partial}%
  \BibitemOpen
  \bibfield  {author} {\bibinfo {author} {\bibfnamefont {L.~A.}\ \bibnamefont
  {Baccal{\'a}}}\ and\ \bibinfo {author} {\bibfnamefont {K.}~\bibnamefont
  {Sameshima}},\ }\bibfield  {title} {\bibinfo {title} {Partial directed
  coherence: a new concept in neural structure determination},\ }\href@noop {}
  {\bibfield  {journal} {\bibinfo  {journal} {Biological cybernetics}\ }\textbf
  {\bibinfo {volume} {84}},\ \bibinfo {pages} {463} (\bibinfo {year}
  {2001})}\BibitemShut {NoStop}%
\bibitem [{\citenamefont {Kami{\'n}ski}\ \emph {et~al.}(2001)\citenamefont
  {Kami{\'n}ski}, \citenamefont {Ding}, \citenamefont {Truccolo},\ and\
  \citenamefont {Bressler}}]{kaminski2001evaluating}%
  \BibitemOpen
  \bibfield  {author} {\bibinfo {author} {\bibfnamefont {M.}~\bibnamefont
  {Kami{\'n}ski}}, \bibinfo {author} {\bibfnamefont {M.}~\bibnamefont {Ding}},
  \bibinfo {author} {\bibfnamefont {W.~A.}\ \bibnamefont {Truccolo}},\ and\
  \bibinfo {author} {\bibfnamefont {S.~L.}\ \bibnamefont {Bressler}},\
  }\bibfield  {title} {\bibinfo {title} {Evaluating causal relations in neural
  systems: Granger causality, directed transfer function and statistical
  assessment of significance},\ }\href@noop {} {\bibfield  {journal} {\bibinfo
  {journal} {Biological cybernetics}\ }\textbf {\bibinfo {volume} {85}},\
  \bibinfo {pages} {145} (\bibinfo {year} {2001})}\BibitemShut {NoStop}%
\bibitem [{\citenamefont {Korzeniewska}\ \emph {et~al.}(2003)\citenamefont
  {Korzeniewska}, \citenamefont {Ma{\'n}czak}, \citenamefont {Kami{\'n}ski},
  \citenamefont {Blinowska},\ and\ \citenamefont
  {Kasicki}}]{korzeniewska2003determination}%
  \BibitemOpen
  \bibfield  {author} {\bibinfo {author} {\bibfnamefont {A.}~\bibnamefont
  {Korzeniewska}}, \bibinfo {author} {\bibfnamefont {M.}~\bibnamefont
  {Ma{\'n}czak}}, \bibinfo {author} {\bibfnamefont {M.}~\bibnamefont
  {Kami{\'n}ski}}, \bibinfo {author} {\bibfnamefont {K.~J.}\ \bibnamefont
  {Blinowska}},\ and\ \bibinfo {author} {\bibfnamefont {S.}~\bibnamefont
  {Kasicki}},\ }\bibfield  {title} {\bibinfo {title} {Determination of
  information flow direction among brain structures by a modified directed
  transfer function (d{DTF}) method},\ }\href@noop {} {\bibfield  {journal}
  {\bibinfo  {journal} {Journal of neuroscience methods}\ }\textbf {\bibinfo
  {volume} {125}},\ \bibinfo {pages} {195} (\bibinfo {year}
  {2003})}\BibitemShut {NoStop}%
\bibitem [{\citenamefont {Schreiber}(2000)}]{schreiber}%
  \BibitemOpen
  \bibfield  {author} {\bibinfo {author} {\bibfnamefont {T.}~\bibnamefont
  {Schreiber}},\ }\bibfield  {title} {\bibinfo {title} {Measuring information
  transfer},\ }\href@noop {} {\bibfield  {journal} {\bibinfo  {journal}
  {Physical Review Letters}\ }\textbf {\bibinfo {volume} {85}},\ \bibinfo
  {pages} {461} (\bibinfo {year} {2000})}\BibitemShut {NoStop}%
\bibitem [{\citenamefont {Palu{\v{s}}}\ \emph {et~al.}(2001)\citenamefont
  {Palu{\v{s}}}, \citenamefont {Kom{\'a}rek}, \citenamefont
  {Hrn{\v{c}}{\'\i}{\v{r}}},\ and\ \citenamefont
  {{\v{S}}t{\v{e}}rbov{\'a}}}]{paluvs2001synchronization}%
  \BibitemOpen
  \bibfield  {author} {\bibinfo {author} {\bibfnamefont {M.}~\bibnamefont
  {Palu{\v{s}}}}, \bibinfo {author} {\bibfnamefont {V.}~\bibnamefont
  {Kom{\'a}rek}}, \bibinfo {author} {\bibfnamefont {Z.}~\bibnamefont
  {Hrn{\v{c}}{\'\i}{\v{r}}}},\ and\ \bibinfo {author} {\bibfnamefont
  {K.}~\bibnamefont {{\v{S}}t{\v{e}}rbov{\'a}}},\ }\bibfield  {title} {\bibinfo
  {title} {Synchronization as adjustment of information rates: Detection from
  bivariate time series},\ }\href@noop {} {\bibfield  {journal} {\bibinfo
  {journal} {Physical Review E}\ }\textbf {\bibinfo {volume} {63}},\ \bibinfo
  {pages} {046211} (\bibinfo {year} {2001})}\BibitemShut {NoStop}%
\bibitem [{\citenamefont {Palu{\v{s}}}\ and\ \citenamefont
  {Vejmelka}(2007)}]{paluvs2007directionality}%
  \BibitemOpen
  \bibfield  {author} {\bibinfo {author} {\bibfnamefont {M.}~\bibnamefont
  {Palu{\v{s}}}}\ and\ \bibinfo {author} {\bibfnamefont {M.}~\bibnamefont
  {Vejmelka}},\ }\bibfield  {title} {\bibinfo {title} {Directionality of
  coupling from bivariate time series: How to avoid false causalities and
  missed connections},\ }\href@noop {} {\bibfield  {journal} {\bibinfo
  {journal} {Physical Review E}\ }\textbf {\bibinfo {volume} {75}},\ \bibinfo
  {pages} {056211} (\bibinfo {year} {2007})}\BibitemShut {NoStop}%
\bibitem [{\citenamefont {Vicente}\ \emph {et~al.}(2011)\citenamefont
  {Vicente}, \citenamefont {Wibral}, \citenamefont {Lindner},\ and\
  \citenamefont {Pipa}}]{vicente2011transfer}%
  \BibitemOpen
  \bibfield  {author} {\bibinfo {author} {\bibfnamefont {R.}~\bibnamefont
  {Vicente}}, \bibinfo {author} {\bibfnamefont {M.}~\bibnamefont {Wibral}},
  \bibinfo {author} {\bibfnamefont {M.}~\bibnamefont {Lindner}},\ and\ \bibinfo
  {author} {\bibfnamefont {G.}~\bibnamefont {Pipa}},\ }\bibfield  {title}
  {\bibinfo {title} {Transfer entropy: a model-free measure of effective
  connectivity for the neurosciences},\ }\href@noop {} {\bibfield  {journal}
  {\bibinfo  {journal} {Journal of computational neuroscience}\ }\textbf
  {\bibinfo {volume} {30}},\ \bibinfo {pages} {45} (\bibinfo {year}
  {2011})}\BibitemShut {NoStop}%
\bibitem [{\citenamefont {Bauer}\ \emph {et~al.}(2007)\citenamefont {Bauer},
  \citenamefont {Cox}, \citenamefont {Caveness}, \citenamefont {Downs},\ and\
  \citenamefont {Thornhill}}]{bauer2007finding}%
  \BibitemOpen
  \bibfield  {author} {\bibinfo {author} {\bibfnamefont {M.}~\bibnamefont
  {Bauer}}, \bibinfo {author} {\bibfnamefont {J.~W.}\ \bibnamefont {Cox}},
  \bibinfo {author} {\bibfnamefont {M.~H.}\ \bibnamefont {Caveness}}, \bibinfo
  {author} {\bibfnamefont {J.~J.}\ \bibnamefont {Downs}},\ and\ \bibinfo
  {author} {\bibfnamefont {N.~F.}\ \bibnamefont {Thornhill}},\ }\bibfield
  {title} {\bibinfo {title} {Finding the direction of disturbance propagation
  in a chemical process using transfer entropy},\ }\href@noop {} {\bibfield
  {journal} {\bibinfo  {journal} {IEEE transactions on control systems
  technology}\ }\textbf {\bibinfo {volume} {15}},\ \bibinfo {pages} {12}
  (\bibinfo {year} {2007})}\BibitemShut {NoStop}%
\bibitem [{\citenamefont {Dimpfl}\ and\ \citenamefont
  {Peter}(2013)}]{dimpfl2013using}%
  \BibitemOpen
  \bibfield  {author} {\bibinfo {author} {\bibfnamefont {T.}~\bibnamefont
  {Dimpfl}}\ and\ \bibinfo {author} {\bibfnamefont {F.~J.}\ \bibnamefont
  {Peter}},\ }\bibfield  {title} {\bibinfo {title} {Using transfer entropy to
  measure information flows between financial markets},\ }\href@noop {}
  {\bibfield  {journal} {\bibinfo  {journal} {Studies in Nonlinear Dynamics \&
  Econometrics}\ }\textbf {\bibinfo {volume} {17}},\ \bibinfo {pages} {85}
  (\bibinfo {year} {2013})}\BibitemShut {NoStop}%
\bibitem [{\citenamefont {Palu{\v{s}}}(2014)}]{paluvs2014multiscale}%
  \BibitemOpen
  \bibfield  {author} {\bibinfo {author} {\bibfnamefont {M.}~\bibnamefont
  {Palu{\v{s}}}},\ }\bibfield  {title} {\bibinfo {title} {Multiscale
  atmospheric dynamics: cross-frequency phase-amplitude coupling in the air
  temperature},\ }\href@noop {} {\bibfield  {journal} {\bibinfo  {journal}
  {Physical review letters}\ }\textbf {\bibinfo {volume} {112}},\ \bibinfo
  {pages} {078702} (\bibinfo {year} {2014})}\BibitemShut {NoStop}%
\bibitem [{\citenamefont {Jajcay}\ \emph {et~al.}(2018)\citenamefont {Jajcay},
  \citenamefont {Kravtsov}, \citenamefont {Sugihara}, \citenamefont {Tsonis},\
  and\ \citenamefont {Palu{\v{s}}}}]{jajcay2018synchronization}%
  \BibitemOpen
  \bibfield  {author} {\bibinfo {author} {\bibfnamefont {N.}~\bibnamefont
  {Jajcay}}, \bibinfo {author} {\bibfnamefont {S.}~\bibnamefont {Kravtsov}},
  \bibinfo {author} {\bibfnamefont {G.}~\bibnamefont {Sugihara}}, \bibinfo
  {author} {\bibfnamefont {A.~A.}\ \bibnamefont {Tsonis}},\ and\ \bibinfo
  {author} {\bibfnamefont {M.}~\bibnamefont {Palu{\v{s}}}},\ }\bibfield
  {title} {\bibinfo {title} {Synchronization and causality across time scales
  in el ni{\~n}o southern oscillation},\ }\href@noop {} {\bibfield  {journal}
  {\bibinfo  {journal} {npj Climate and Atmospheric Science}\ }\textbf
  {\bibinfo {volume} {1}},\ \bibinfo {pages} {1} (\bibinfo {year}
  {2018})}\BibitemShut {NoStop}%
\bibitem [{\citenamefont {Takens}(1981)}]{taken}%
  \BibitemOpen
  \bibfield  {author} {\bibinfo {author} {\bibfnamefont {F.}~\bibnamefont
  {Takens}},\ }\bibfield  {title} {\bibinfo {title} {Detecting strange
  attractors in turbulence},\ }in\ \href@noop {} {\emph {\bibinfo {booktitle}
  {Dynamical systems and turbulence, Warwick 1980}}}\ (\bibinfo  {publisher}
  {Springer},\ \bibinfo {year} {1981})\ pp.\ \bibinfo {pages}
  {366--381}\BibitemShut {NoStop}%
\bibitem [{\citenamefont {Fraser}\ and\ \citenamefont
  {Swinney}(1986)}]{fraser1986independent}%
  \BibitemOpen
  \bibfield  {author} {\bibinfo {author} {\bibfnamefont {A.~M.}\ \bibnamefont
  {Fraser}}\ and\ \bibinfo {author} {\bibfnamefont {H.~L.}\ \bibnamefont
  {Swinney}},\ }\bibfield  {title} {\bibinfo {title} {Independent coordinates
  for strange attractors from mutual information},\ }\href@noop {} {\bibfield
  {journal} {\bibinfo  {journal} {Physical review A}\ }\textbf {\bibinfo
  {volume} {33}},\ \bibinfo {pages} {1134} (\bibinfo {year}
  {1986})}\BibitemShut {NoStop}%
\bibitem [{\citenamefont {Wibral}\ \emph {et~al.}(2013)\citenamefont {Wibral},
  \citenamefont {Pampu}, \citenamefont {Priesemann}, \citenamefont
  {Siebenh{\"u}hner}, \citenamefont {Seiwert}, \citenamefont {Lindner},
  \citenamefont {Lizier},\ and\ \citenamefont {Vicente}}]{wibral2013measuring}%
  \BibitemOpen
  \bibfield  {author} {\bibinfo {author} {\bibfnamefont {M.}~\bibnamefont
  {Wibral}}, \bibinfo {author} {\bibfnamefont {N.}~\bibnamefont {Pampu}},
  \bibinfo {author} {\bibfnamefont {V.}~\bibnamefont {Priesemann}}, \bibinfo
  {author} {\bibfnamefont {F.}~\bibnamefont {Siebenh{\"u}hner}}, \bibinfo
  {author} {\bibfnamefont {H.}~\bibnamefont {Seiwert}}, \bibinfo {author}
  {\bibfnamefont {M.}~\bibnamefont {Lindner}}, \bibinfo {author} {\bibfnamefont
  {J.~T.}\ \bibnamefont {Lizier}},\ and\ \bibinfo {author} {\bibfnamefont
  {R.}~\bibnamefont {Vicente}},\ }\bibfield  {title} {\bibinfo {title}
  {Measuring information-transfer delays},\ }\href@noop {} {\bibfield
  {journal} {\bibinfo  {journal} {PloS one}\ }\textbf {\bibinfo {volume} {8}},\
  \bibinfo {pages} {e55809} (\bibinfo {year} {2013})}\BibitemShut {NoStop}%
\bibitem [{\citenamefont {Sugihara}\ \emph {et~al.}(2012)\citenamefont
  {Sugihara}, \citenamefont {May}, \citenamefont {Ye}, \citenamefont {Hsieh},\
  and\ \citenamefont {Deyle}}]{sugihara}%
  \BibitemOpen
  \bibfield  {author} {\bibinfo {author} {\bibfnamefont {G.}~\bibnamefont
  {Sugihara}}, \bibinfo {author} {\bibfnamefont {R.}~\bibnamefont {May}},
  \bibinfo {author} {\bibfnamefont {H.}~\bibnamefont {Ye}}, \bibinfo {author}
  {\bibfnamefont {C.}~\bibnamefont {Hsieh}},\ and\ \bibinfo {author}
  {\bibfnamefont {E.}~\bibnamefont {Deyle}},\ }\bibfield  {title} {\bibinfo
  {title} {Detecting causality in complex ecosystems},\ }\href@noop {}
  {\bibfield  {journal} {\bibinfo  {journal} {Science}\ }\textbf {\bibinfo
  {volume} {338}},\ \bibinfo {pages} {496} (\bibinfo {year}
  {2012})}\BibitemShut {NoStop}%
\bibitem [{\citenamefont {Harnack}\ \emph {et~al.}(2017)\citenamefont
  {Harnack}, \citenamefont {Laminski}, \citenamefont {Sch{\"u}nemann},\ and\
  \citenamefont {Pawelzik}}]{harnack2017topological}%
  \BibitemOpen
  \bibfield  {author} {\bibinfo {author} {\bibfnamefont {D.}~\bibnamefont
  {Harnack}}, \bibinfo {author} {\bibfnamefont {E.}~\bibnamefont {Laminski}},
  \bibinfo {author} {\bibfnamefont {M.}~\bibnamefont {Sch{\"u}nemann}},\ and\
  \bibinfo {author} {\bibfnamefont {K.~R.}\ \bibnamefont {Pawelzik}},\
  }\bibfield  {title} {\bibinfo {title} {Topological causality in dynamical
  systems},\ }\href@noop {} {\bibfield  {journal} {\bibinfo  {journal}
  {Physical review letters}\ }\textbf {\bibinfo {volume} {119}},\ \bibinfo
  {pages} {098301} (\bibinfo {year} {2017})}\BibitemShut {NoStop}%
\bibitem [{\citenamefont {Krakovsk{\'a}}\ and\ \citenamefont
  {Hanzely}(2016)}]{krakovska2016testing}%
  \BibitemOpen
  \bibfield  {author} {\bibinfo {author} {\bibfnamefont {A.}~\bibnamefont
  {Krakovsk{\'a}}}\ and\ \bibinfo {author} {\bibfnamefont {F.}~\bibnamefont
  {Hanzely}},\ }\bibfield  {title} {\bibinfo {title} {Testing for causality in
  reconstructed state spaces by an optimized mixed prediction method},\
  }\href@noop {} {\bibfield  {journal} {\bibinfo  {journal} {Physical Review
  E}\ }\textbf {\bibinfo {volume} {94}},\ \bibinfo {pages} {052203} (\bibinfo
  {year} {2016})}\BibitemShut {NoStop}%
\bibitem [{\citenamefont {Barrios}\ \emph {et~al.}(2018)\citenamefont
  {Barrios}, \citenamefont {Trincado},\ and\ \citenamefont
  {Garreaud}}]{barrios2018alternative}%
  \BibitemOpen
  \bibfield  {author} {\bibinfo {author} {\bibfnamefont {A.}~\bibnamefont
  {Barrios}}, \bibinfo {author} {\bibfnamefont {G.}~\bibnamefont {Trincado}},\
  and\ \bibinfo {author} {\bibfnamefont {R.}~\bibnamefont {Garreaud}},\
  }\bibfield  {title} {\bibinfo {title} {Alternative approaches for estimating
  missing climate data: application to monthly precipitation records in
  south-central chile},\ }\href@noop {} {\bibfield  {journal} {\bibinfo
  {journal} {Forest Ecosystems}\ }\textbf {\bibinfo {volume} {5}},\ \bibinfo
  {pages} {1} (\bibinfo {year} {2018})}\BibitemShut {NoStop}%
\bibitem [{\citenamefont {Anderson}\ and\ \citenamefont
  {Gough}(2018)}]{anderson2018accounting}%
  \BibitemOpen
  \bibfield  {author} {\bibinfo {author} {\bibfnamefont {C.~I.}\ \bibnamefont
  {Anderson}}\ and\ \bibinfo {author} {\bibfnamefont {W.~A.}\ \bibnamefont
  {Gough}},\ }\bibfield  {title} {\bibinfo {title} {Accounting for missing data
  in monthly temperature series: Testing rule-of-thumb omission of months with
  missing values},\ }\href@noop {} {\bibfield  {journal} {\bibinfo  {journal}
  {International Journal of Climatology}\ }\textbf {\bibinfo {volume} {38}},\
  \bibinfo {pages} {4990} (\bibinfo {year} {2018})}\BibitemShut {NoStop}%
\bibitem [{\citenamefont {DiCesare}(2006)}]{dicesare2006imputation}%
  \BibitemOpen
  \bibfield  {author} {\bibinfo {author} {\bibfnamefont {G.}~\bibnamefont
  {DiCesare}},\ }\emph {\bibinfo {title} {Imputation, estimation and missing
  data in finance}},\ \href@noop {} {Ph.D. thesis},\ \bibinfo  {school}
  {University of Waterloo} (\bibinfo {year} {2006})\BibitemShut {NoStop}%
\bibitem [{\citenamefont {John}\ \emph {et~al.}(2019)\citenamefont {John},
  \citenamefont {Ekpenyong},\ and\ \citenamefont {Nworu}}]{john2019imputation}%
  \BibitemOpen
  \bibfield  {author} {\bibinfo {author} {\bibfnamefont {C.}~\bibnamefont
  {John}}, \bibinfo {author} {\bibfnamefont {E.~J.}\ \bibnamefont
  {Ekpenyong}},\ and\ \bibinfo {author} {\bibfnamefont {C.~C.}\ \bibnamefont
  {Nworu}},\ }\bibfield  {title} {\bibinfo {title} {Imputation of missing
  values in economic and financial time series data using five principal
  component analysis approaches},\ }\href@noop {} {\bibfield  {journal}
  {\bibinfo  {journal} {CBN Journal of Applied Statistics (JAS)}\ }\textbf
  {\bibinfo {volume} {10}},\ \bibinfo {pages} {3} (\bibinfo {year}
  {2019})}\BibitemShut {NoStop}%
\bibitem [{\citenamefont {Gyimah}(2001)}]{gyimah2001missing}%
  \BibitemOpen
  \bibfield  {author} {\bibinfo {author} {\bibfnamefont {S.}~\bibnamefont
  {Gyimah}},\ }\bibfield  {title} {\bibinfo {title} {Missing data in
  quantitative social research},\ }\href@noop {} {\bibfield  {journal}
  {\bibinfo  {journal} {PSC Discussion Papers Series}\ }\textbf {\bibinfo
  {volume} {15}},\ \bibinfo {pages} {1} (\bibinfo {year} {2001})}\BibitemShut
  {NoStop}%
\bibitem [{\citenamefont {Kulp}\ and\ \citenamefont
  {Tracy}(2009)}]{kulp2009application}%
  \BibitemOpen
  \bibfield  {author} {\bibinfo {author} {\bibfnamefont {C.}~\bibnamefont
  {Kulp}}\ and\ \bibinfo {author} {\bibfnamefont {E.}~\bibnamefont {Tracy}},\
  }\bibfield  {title} {\bibinfo {title} {The application of the transfer
  entropy to gappy time series},\ }\href@noop {} {\bibfield  {journal}
  {\bibinfo  {journal} {Physics Letters A}\ }\textbf {\bibinfo {volume}
  {373}},\ \bibinfo {pages} {1261} (\bibinfo {year} {2009})}\BibitemShut
  {NoStop}%
\bibitem [{\citenamefont {Smirnov}\ and\ \citenamefont
  {Bezruchko}(2012)}]{smirnov2012spurious}%
  \BibitemOpen
  \bibfield  {author} {\bibinfo {author} {\bibfnamefont {D.}~\bibnamefont
  {Smirnov}}\ and\ \bibinfo {author} {\bibfnamefont {B.}~\bibnamefont
  {Bezruchko}},\ }\bibfield  {title} {\bibinfo {title} {Spurious causalities
  due to low temporal resolution: Towards detection of bidirectional coupling
  from time series},\ }\href@noop {} {\bibfield  {journal} {\bibinfo  {journal}
  {EPL (Europhysics Letters)}\ }\textbf {\bibinfo {volume} {100}},\ \bibinfo
  {pages} {10005} (\bibinfo {year} {2012})}\BibitemShut {NoStop}%
\bibitem [{\citenamefont {Kathpalia}\ and\ \citenamefont
  {Nagaraj}(2019)}]{kathpalia2019data}%
  \BibitemOpen
  \bibfield  {author} {\bibinfo {author} {\bibfnamefont {A.}~\bibnamefont
  {Kathpalia}}\ and\ \bibinfo {author} {\bibfnamefont {N.}~\bibnamefont
  {Nagaraj}},\ }\bibfield  {title} {\bibinfo {title} {Data based intervention
  approach for complexity-causality measure},\ }\href@noop {} {\bibfield
  {journal} {\bibinfo  {journal} {PeerJ Computer Science}\ }\textbf {\bibinfo
  {volume} {5}} (\bibinfo {year} {2019})}\BibitemShut {NoStop}%
\bibitem [{\citenamefont {Kathpalia}(2021)}]{kathpalia2021theoretical}%
  \BibitemOpen
  \bibfield  {author} {\bibinfo {author} {\bibfnamefont {A.}~\bibnamefont
  {Kathpalia}},\ }\emph {\bibinfo {title} {Theoretical and Experimental
  Investigations into Causality, its Measures and Applications}},\ \href@noop
  {} {Ph.D. thesis},\ \bibinfo  {school} {NIAS} (\bibinfo {year}
  {2021})\BibitemShut {NoStop}%
\bibitem [{\citenamefont {Nagaraj}\ and\ \citenamefont
  {Balasubramanian}(2017)}]{nagaraj2}%
  \BibitemOpen
  \bibfield  {author} {\bibinfo {author} {\bibfnamefont {N.}~\bibnamefont
  {Nagaraj}}\ and\ \bibinfo {author} {\bibfnamefont {K.}~\bibnamefont
  {Balasubramanian}},\ }\bibfield  {title} {\bibinfo {title} {Dynamical
  complexity of short and noisy time series},\ }\href@noop {} {\bibfield
  {journal} {\bibinfo  {journal} {The European Physical Journal Special
  Topics}\ ,\ \bibinfo {pages} {1}} (\bibinfo {year} {2017})}\BibitemShut
  {NoStop}%
\bibitem [{\citenamefont {Staniek}\ and\ \citenamefont
  {Lehnertz}(2008)}]{staniek2008symbolic}%
  \BibitemOpen
  \bibfield  {author} {\bibinfo {author} {\bibfnamefont {M.}~\bibnamefont
  {Staniek}}\ and\ \bibinfo {author} {\bibfnamefont {K.}~\bibnamefont
  {Lehnertz}},\ }\bibfield  {title} {\bibinfo {title} {Symbolic transfer
  entropy},\ }\href@noop {} {\bibfield  {journal} {\bibinfo  {journal}
  {Physical Review Letters}\ }\textbf {\bibinfo {volume} {100}},\ \bibinfo
  {pages} {158101} (\bibinfo {year} {2008})}\BibitemShut {NoStop}%
\bibitem [{\citenamefont {Staniek}\ and\ \citenamefont
  {Lehnertz}(2009)}]{staniek2009symbolic}%
  \BibitemOpen
  \bibfield  {author} {\bibinfo {author} {\bibfnamefont {M.}~\bibnamefont
  {Staniek}}\ and\ \bibinfo {author} {\bibfnamefont {K.}~\bibnamefont
  {Lehnertz}},\ }\bibfield  {title} {\bibinfo {title} {Symbolic transfer
  entropy: inferring directionality in biosignals},\ }\href@noop {} {\bibfield
  {journal} {\bibinfo  {journal} {Biomedizinische Technik}\ }\textbf {\bibinfo
  {volume} {54}},\ \bibinfo {pages} {323} (\bibinfo {year} {2009})}\BibitemShut
  {NoStop}%
\bibitem [{\citenamefont {Kugiumtzis}(2013)}]{kugiumtzis2013partial}%
  \BibitemOpen
  \bibfield  {author} {\bibinfo {author} {\bibfnamefont {D.}~\bibnamefont
  {Kugiumtzis}},\ }\bibfield  {title} {\bibinfo {title} {Partial transfer
  entropy on rank vectors},\ }\href@noop {} {\bibfield  {journal} {\bibinfo
  {journal} {The European Physical Journal Special Topics}\ }\textbf {\bibinfo
  {volume} {222}},\ \bibinfo {pages} {401} (\bibinfo {year}
  {2013})}\BibitemShut {NoStop}%
\bibitem [{\citenamefont {Papana}\ \emph {et~al.}(2013)\citenamefont {Papana},
  \citenamefont {Kyrtsou}, \citenamefont {Kugiumtzis},\ and\ \citenamefont
  {Diks}}]{papana2013simulation}%
  \BibitemOpen
  \bibfield  {author} {\bibinfo {author} {\bibfnamefont {A.}~\bibnamefont
  {Papana}}, \bibinfo {author} {\bibfnamefont {C.}~\bibnamefont {Kyrtsou}},
  \bibinfo {author} {\bibfnamefont {D.}~\bibnamefont {Kugiumtzis}},\ and\
  \bibinfo {author} {\bibfnamefont {C.}~\bibnamefont {Diks}},\ }\bibfield
  {title} {\bibinfo {title} {Simulation study of direct causality measures in
  multivariate time series},\ }\href@noop {} {\bibfield  {journal} {\bibinfo
  {journal} {Entropy}\ }\textbf {\bibinfo {volume} {15}},\ \bibinfo {pages}
  {2635} (\bibinfo {year} {2013})}\BibitemShut {NoStop}%
\bibitem [{\citenamefont {Li}\ and\ \citenamefont
  {Ouyang}(2010)}]{li2010estimating}%
  \BibitemOpen
  \bibfield  {author} {\bibinfo {author} {\bibfnamefont {X.}~\bibnamefont
  {Li}}\ and\ \bibinfo {author} {\bibfnamefont {G.}~\bibnamefont {Ouyang}},\
  }\bibfield  {title} {\bibinfo {title} {Estimating coupling direction between
  neuronal populations with permutation conditional mutual information},\
  }\href@noop {} {\bibfield  {journal} {\bibinfo  {journal} {NeuroImage}\
  }\textbf {\bibinfo {volume} {52}},\ \bibinfo {pages} {497} (\bibinfo {year}
  {2010})}\BibitemShut {NoStop}%
\bibitem [{\citenamefont {Wen}\ \emph {et~al.}(2019)\citenamefont {Wen},
  \citenamefont {Jia}, \citenamefont {Hsu}, \citenamefont {Zhou}, \citenamefont
  {Lan}, \citenamefont {Cui}, \citenamefont {Li}, \citenamefont {Yin},\ and\
  \citenamefont {Wang}}]{wen2019estimating}%
  \BibitemOpen
  \bibfield  {author} {\bibinfo {author} {\bibfnamefont {D.}~\bibnamefont
  {Wen}}, \bibinfo {author} {\bibfnamefont {P.}~\bibnamefont {Jia}}, \bibinfo
  {author} {\bibfnamefont {S.-H.}\ \bibnamefont {Hsu}}, \bibinfo {author}
  {\bibfnamefont {Y.}~\bibnamefont {Zhou}}, \bibinfo {author} {\bibfnamefont
  {X.}~\bibnamefont {Lan}}, \bibinfo {author} {\bibfnamefont {D.}~\bibnamefont
  {Cui}}, \bibinfo {author} {\bibfnamefont {G.}~\bibnamefont {Li}}, \bibinfo
  {author} {\bibfnamefont {S.}~\bibnamefont {Yin}},\ and\ \bibinfo {author}
  {\bibfnamefont {L.}~\bibnamefont {Wang}},\ }\bibfield  {title} {\bibinfo
  {title} {Estimating coupling strength between multivariate neural series with
  multivariate permutation conditional mutual information},\ }\href@noop {}
  {\bibfield  {journal} {\bibinfo  {journal} {Neural Networks}\ }\textbf
  {\bibinfo {volume} {110}},\ \bibinfo {pages} {159} (\bibinfo {year}
  {2019})}\BibitemShut {NoStop}%
\bibitem [{\citenamefont {Bandt}\ and\ \citenamefont
  {Pompe}(2002)}]{bandt2002permutation}%
  \BibitemOpen
  \bibfield  {author} {\bibinfo {author} {\bibfnamefont {C.}~\bibnamefont
  {Bandt}}\ and\ \bibinfo {author} {\bibfnamefont {B.}~\bibnamefont {Pompe}},\
  }\bibfield  {title} {\bibinfo {title} {Permutation entropy: a natural
  complexity measure for time series},\ }\href@noop {} {\bibfield  {journal}
  {\bibinfo  {journal} {Physical review letters}\ }\textbf {\bibinfo {volume}
  {88}},\ \bibinfo {pages} {174102} (\bibinfo {year} {2002})}\BibitemShut
  {NoStop}%
\bibitem [{\citenamefont {Fadlallah}\ \emph {et~al.}(2013)\citenamefont
  {Fadlallah}, \citenamefont {Chen}, \citenamefont {Keil},\ and\ \citenamefont
  {Principe}}]{fadlallah2013weighted}%
  \BibitemOpen
  \bibfield  {author} {\bibinfo {author} {\bibfnamefont {B.}~\bibnamefont
  {Fadlallah}}, \bibinfo {author} {\bibfnamefont {B.}~\bibnamefont {Chen}},
  \bibinfo {author} {\bibfnamefont {A.}~\bibnamefont {Keil}},\ and\ \bibinfo
  {author} {\bibfnamefont {J.}~\bibnamefont {Principe}},\ }\bibfield  {title}
  {\bibinfo {title} {Weighted-permutation entropy: A complexity measure for
  time series incorporating amplitude information},\ }\href@noop {} {\bibfield
  {journal} {\bibinfo  {journal} {Physical Review E}\ }\textbf {\bibinfo
  {volume} {87}},\ \bibinfo {pages} {022911} (\bibinfo {year}
  {2013})}\BibitemShut {NoStop}%
\bibitem [{\citenamefont {Amig{\'o}}(2010)}]{amigo2010permutation}%
  \BibitemOpen
  \bibfield  {author} {\bibinfo {author} {\bibfnamefont {J.}~\bibnamefont
  {Amig{\'o}}},\ }\href@noop {} {\emph {\bibinfo {title} {Permutation
  complexity in dynamical systems: ordinal patterns, permutation entropy and
  all that}}}\ (\bibinfo  {publisher} {Springer Science \& Business Media},\
  \bibinfo {year} {2010})\BibitemShut {NoStop}%
\bibitem [{\citenamefont {Zanin}\ \emph {et~al.}(2012)\citenamefont {Zanin},
  \citenamefont {Zunino}, \citenamefont {Rosso},\ and\ \citenamefont
  {Papo}}]{zanin2012permutation}%
  \BibitemOpen
  \bibfield  {author} {\bibinfo {author} {\bibfnamefont {M.}~\bibnamefont
  {Zanin}}, \bibinfo {author} {\bibfnamefont {L.}~\bibnamefont {Zunino}},
  \bibinfo {author} {\bibfnamefont {O.~A.}\ \bibnamefont {Rosso}},\ and\
  \bibinfo {author} {\bibfnamefont {D.}~\bibnamefont {Papo}},\ }\bibfield
  {title} {\bibinfo {title} {Permutation entropy and its main biomedical and
  econophysics applications: a review},\ }\href@noop {} {\bibfield  {journal}
  {\bibinfo  {journal} {Entropy}\ }\textbf {\bibinfo {volume} {14}},\ \bibinfo
  {pages} {1553} (\bibinfo {year} {2012})}\BibitemShut {NoStop}%
\bibitem [{\citenamefont {Keller}\ \emph {et~al.}(2014)\citenamefont {Keller},
  \citenamefont {Unakafov},\ and\ \citenamefont
  {Unakafova}}]{keller2014ordinal}%
  \BibitemOpen
  \bibfield  {author} {\bibinfo {author} {\bibfnamefont {K.}~\bibnamefont
  {Keller}}, \bibinfo {author} {\bibfnamefont {A.~M.}\ \bibnamefont
  {Unakafov}},\ and\ \bibinfo {author} {\bibfnamefont {V.~A.}\ \bibnamefont
  {Unakafova}},\ }\bibfield  {title} {\bibinfo {title} {Ordinal patterns,
  entropy, and eeg},\ }\href@noop {} {\bibfield  {journal} {\bibinfo  {journal}
  {Entropy}\ }\textbf {\bibinfo {volume} {16}},\ \bibinfo {pages} {6212}
  (\bibinfo {year} {2014})}\BibitemShut {NoStop}%
\bibitem [{\citenamefont {Zanin}\ and\ \citenamefont
  {Olivares}(2021)}]{zanin2021ordinal}%
  \BibitemOpen
  \bibfield  {author} {\bibinfo {author} {\bibfnamefont {M.}~\bibnamefont
  {Zanin}}\ and\ \bibinfo {author} {\bibfnamefont {F.}~\bibnamefont
  {Olivares}},\ }\bibfield  {title} {\bibinfo {title} {Ordinal patterns-based
  methodologies for distinguishing chaos from noise in discrete time series},\
  }\href@noop {} {\bibfield  {journal} {\bibinfo  {journal} {Communications
  Physics}\ }\textbf {\bibinfo {volume} {4}},\ \bibinfo {pages} {1} (\bibinfo
  {year} {2021})}\BibitemShut {NoStop}%
\bibitem [{\citenamefont {McCullough}\ \emph {et~al.}(2015)\citenamefont
  {McCullough}, \citenamefont {Small}, \citenamefont {Stemler},\ and\
  \citenamefont {Iu}}]{mccullough2015time}%
  \BibitemOpen
  \bibfield  {author} {\bibinfo {author} {\bibfnamefont {M.}~\bibnamefont
  {McCullough}}, \bibinfo {author} {\bibfnamefont {M.}~\bibnamefont {Small}},
  \bibinfo {author} {\bibfnamefont {T.}~\bibnamefont {Stemler}},\ and\ \bibinfo
  {author} {\bibfnamefont {H.~H.-C.}\ \bibnamefont {Iu}},\ }\bibfield  {title}
  {\bibinfo {title} {Time lagged ordinal partition networks for capturing
  dynamics of continuous dynamical systems},\ }\href@noop {} {\bibfield
  {journal} {\bibinfo  {journal} {Chaos: An Interdisciplinary Journal of
  Nonlinear Science}\ }\textbf {\bibinfo {volume} {25}},\ \bibinfo {pages}
  {053101} (\bibinfo {year} {2015})}\BibitemShut {NoStop}%
\bibitem [{\citenamefont {Bandt}\ \emph {et~al.}(2002)\citenamefont {Bandt},
  \citenamefont {Keller},\ and\ \citenamefont {Pompe}}]{bandt2002entropy}%
  \BibitemOpen
  \bibfield  {author} {\bibinfo {author} {\bibfnamefont {C.}~\bibnamefont
  {Bandt}}, \bibinfo {author} {\bibfnamefont {G.}~\bibnamefont {Keller}},\ and\
  \bibinfo {author} {\bibfnamefont {B.}~\bibnamefont {Pompe}},\ }\bibfield
  {title} {\bibinfo {title} {Entropy of interval maps via permutations},\
  }\href@noop {} {\bibfield  {journal} {\bibinfo  {journal} {Nonlinearity}\
  }\textbf {\bibinfo {volume} {15}},\ \bibinfo {pages} {1595} (\bibinfo {year}
  {2002})}\BibitemShut {NoStop}%
\bibitem [{\citenamefont {Amig{\'o}}\ \emph {et~al.}(2005)\citenamefont
  {Amig{\'o}}, \citenamefont {Kennel},\ and\ \citenamefont
  {Kocarev}}]{amigo2005permutation}%
  \BibitemOpen
  \bibfield  {author} {\bibinfo {author} {\bibfnamefont {J.~M.}\ \bibnamefont
  {Amig{\'o}}}, \bibinfo {author} {\bibfnamefont {M.~B.}\ \bibnamefont
  {Kennel}},\ and\ \bibinfo {author} {\bibfnamefont {L.}~\bibnamefont
  {Kocarev}},\ }\bibfield  {title} {\bibinfo {title} {The permutation entropy
  rate equals the metric entropy rate for ergodic information sources and
  ergodic dynamical systems},\ }\href@noop {} {\bibfield  {journal} {\bibinfo
  {journal} {Physica D: Nonlinear Phenomena}\ }\textbf {\bibinfo {volume}
  {210}},\ \bibinfo {pages} {77} (\bibinfo {year} {2005})}\BibitemShut
  {NoStop}%
\bibitem [{\citenamefont {Solomon}\ \emph {et~al.}(2007)\citenamefont
  {Solomon}, \citenamefont {Manning}, \citenamefont {Marquis}, \citenamefont
  {Qin} \emph {et~al.}}]{solomon2007climate}%
  \BibitemOpen
  \bibfield  {author} {\bibinfo {author} {\bibfnamefont {S.}~\bibnamefont
  {Solomon}}, \bibinfo {author} {\bibfnamefont {M.}~\bibnamefont {Manning}},
  \bibinfo {author} {\bibfnamefont {M.}~\bibnamefont {Marquis}}, \bibinfo
  {author} {\bibfnamefont {D.}~\bibnamefont {Qin}}, \emph {et~al.},\
  }\href@noop {} {\emph {\bibinfo {title} {Climate change 2007-the physical
  science basis: Working group I contribution to the fourth assessment report
  of the IPCC}}},\ Vol.~\bibinfo {volume} {4}\ (\bibinfo  {publisher}
  {Cambridge university press},\ \bibinfo {year} {2007})\BibitemShut {NoStop}%
\bibitem [{\citenamefont {Press}\ \emph {et~al.}(1987)\citenamefont {Press},
  \citenamefont {Flannery}, \citenamefont {Teukolsky}, \citenamefont
  {Vetterling},\ and\ \citenamefont {Kramer}}]{press1987numerical}%
  \BibitemOpen
  \bibfield  {author} {\bibinfo {author} {\bibfnamefont {W.~H.}\ \bibnamefont
  {Press}}, \bibinfo {author} {\bibfnamefont {B.~P.}\ \bibnamefont {Flannery}},
  \bibinfo {author} {\bibfnamefont {S.~A.}\ \bibnamefont {Teukolsky}}, \bibinfo
  {author} {\bibfnamefont {W.~T.}\ \bibnamefont {Vetterling}},\ and\ \bibinfo
  {author} {\bibfnamefont {P.~B.}\ \bibnamefont {Kramer}},\ }\bibfield  {title}
  {\bibinfo {title} {Numerical recipes: the art of scientific computing},\
  }\href@noop {} {\bibfield  {journal} {\bibinfo  {journal} {Physics Today}\
  }\textbf {\bibinfo {volume} {40}},\ \bibinfo {pages} {120} (\bibinfo {year}
  {1987})}\BibitemShut {NoStop}%
\bibitem [{\citenamefont {Theiler}\ \emph {et~al.}(1992)\citenamefont
  {Theiler}, \citenamefont {Eubank}, \citenamefont {Longtin}, \citenamefont
  {Galdrikian},\ and\ \citenamefont {Farmer}}]{theiler1992testing}%
  \BibitemOpen
  \bibfield  {author} {\bibinfo {author} {\bibfnamefont {J.}~\bibnamefont
  {Theiler}}, \bibinfo {author} {\bibfnamefont {S.}~\bibnamefont {Eubank}},
  \bibinfo {author} {\bibfnamefont {A.}~\bibnamefont {Longtin}}, \bibinfo
  {author} {\bibfnamefont {B.}~\bibnamefont {Galdrikian}},\ and\ \bibinfo
  {author} {\bibfnamefont {J.~D.}\ \bibnamefont {Farmer}},\ }\bibfield  {title}
  {\bibinfo {title} {Testing for nonlinearity in time series: the method of
  surrogate data},\ }\href@noop {} {\bibfield  {journal} {\bibinfo  {journal}
  {Physica D: Nonlinear Phenomena}\ }\textbf {\bibinfo {volume} {58}},\
  \bibinfo {pages} {77} (\bibinfo {year} {1992})}\BibitemShut {NoStop}%
\bibitem [{\citenamefont {Mills}\ \emph {et~al.}(2019)\citenamefont {Mills},
  \citenamefont {Krause}, \citenamefont {Scotese}, \citenamefont {Hill},
  \citenamefont {Shields},\ and\ \citenamefont {Lenton}}]{mills2019modelling}%
  \BibitemOpen
  \bibfield  {author} {\bibinfo {author} {\bibfnamefont {B.~J.}\ \bibnamefont
  {Mills}}, \bibinfo {author} {\bibfnamefont {A.~J.}\ \bibnamefont {Krause}},
  \bibinfo {author} {\bibfnamefont {C.~R.}\ \bibnamefont {Scotese}}, \bibinfo
  {author} {\bibfnamefont {D.~J.}\ \bibnamefont {Hill}}, \bibinfo {author}
  {\bibfnamefont {G.~A.}\ \bibnamefont {Shields}},\ and\ \bibinfo {author}
  {\bibfnamefont {T.~M.}\ \bibnamefont {Lenton}},\ }\bibfield  {title}
  {\bibinfo {title} {Modelling the long-term carbon cycle, atmospheric co2, and
  earth surface temperature from late neoproterozoic to present day},\
  }\href@noop {} {\bibfield  {journal} {\bibinfo  {journal} {Gondwana
  Research}\ }\textbf {\bibinfo {volume} {67}},\ \bibinfo {pages} {172}
  (\bibinfo {year} {2019})}\BibitemShut {NoStop}%
\bibitem [{\citenamefont {Wong}\ \emph {et~al.}(2021)\citenamefont {Wong},
  \citenamefont {Cui}, \citenamefont {Royer},\ and\ \citenamefont
  {Keller}}]{wong2021tighter}%
  \BibitemOpen
  \bibfield  {author} {\bibinfo {author} {\bibfnamefont {T.~E.}\ \bibnamefont
  {Wong}}, \bibinfo {author} {\bibfnamefont {Y.}~\bibnamefont {Cui}}, \bibinfo
  {author} {\bibfnamefont {D.~L.}\ \bibnamefont {Royer}},\ and\ \bibinfo
  {author} {\bibfnamefont {K.}~\bibnamefont {Keller}},\ }\bibfield  {title}
  {\bibinfo {title} {A tighter constraint on earth-system sensitivity from
  long-term temperature and carbon-cycle observations},\ }\href@noop {}
  {\bibfield  {journal} {\bibinfo  {journal} {Nature communications}\ }\textbf
  {\bibinfo {volume} {12}},\ \bibinfo {pages} {1} (\bibinfo {year}
  {2021})}\BibitemShut {NoStop}%
\bibitem [{\citenamefont {of~PAGES}(2016)}]{past2016interglacials}%
  \BibitemOpen
  \bibfield  {author} {\bibinfo {author} {\bibfnamefont {P.~I. W.~G.}\
  \bibnamefont {of~PAGES}},\ }\bibfield  {title} {\bibinfo {title}
  {Interglacials of the last 800,000 years},\ }\href@noop {} {\bibfield
  {journal} {\bibinfo  {journal} {Reviews of Geophysics}\ }\textbf {\bibinfo
  {volume} {54}},\ \bibinfo {pages} {162} (\bibinfo {year} {2016})}\BibitemShut
  {NoStop}%
\bibitem [{\citenamefont {L{\"u}thi}\ \emph {et~al.}(2008)\citenamefont
  {L{\"u}thi}, \citenamefont {Le~Floch}, \citenamefont {Bereiter},
  \citenamefont {Blunier}, \citenamefont {Barnola}, \citenamefont
  {Siegenthaler}, \citenamefont {Raynaud}, \citenamefont {Jouzel},
  \citenamefont {Fischer}, \citenamefont {Kawamura} \emph
  {et~al.}}]{luthi2008high}%
  \BibitemOpen
  \bibfield  {author} {\bibinfo {author} {\bibfnamefont {D.}~\bibnamefont
  {L{\"u}thi}}, \bibinfo {author} {\bibfnamefont {M.}~\bibnamefont {Le~Floch}},
  \bibinfo {author} {\bibfnamefont {B.}~\bibnamefont {Bereiter}}, \bibinfo
  {author} {\bibfnamefont {T.}~\bibnamefont {Blunier}}, \bibinfo {author}
  {\bibfnamefont {J.-M.}\ \bibnamefont {Barnola}}, \bibinfo {author}
  {\bibfnamefont {U.}~\bibnamefont {Siegenthaler}}, \bibinfo {author}
  {\bibfnamefont {D.}~\bibnamefont {Raynaud}}, \bibinfo {author} {\bibfnamefont
  {J.}~\bibnamefont {Jouzel}}, \bibinfo {author} {\bibfnamefont
  {H.}~\bibnamefont {Fischer}}, \bibinfo {author} {\bibfnamefont
  {K.}~\bibnamefont {Kawamura}}, \emph {et~al.},\ }\bibfield  {title} {\bibinfo
  {title} {High-resolution carbon dioxide concentration record 650,000--800,000
  years before present},\ }\href@noop {} {\bibfield  {journal} {\bibinfo
  {journal} {nature}\ }\textbf {\bibinfo {volume} {453}},\ \bibinfo {pages}
  {379} (\bibinfo {year} {2008})}\BibitemShut {NoStop}%
\bibitem [{\citenamefont {Bereiter}\ \emph {et~al.}(2015)\citenamefont
  {Bereiter}, \citenamefont {Eggleston}, \citenamefont {Schmitt}, \citenamefont
  {Nehrbass-Ahles}, \citenamefont {Stocker}, \citenamefont {Fischer},
  \citenamefont {Kipfstuhl},\ and\ \citenamefont
  {Chappellaz}}]{bereiter2015revision}%
  \BibitemOpen
  \bibfield  {author} {\bibinfo {author} {\bibfnamefont {B.}~\bibnamefont
  {Bereiter}}, \bibinfo {author} {\bibfnamefont {S.}~\bibnamefont {Eggleston}},
  \bibinfo {author} {\bibfnamefont {J.}~\bibnamefont {Schmitt}}, \bibinfo
  {author} {\bibfnamefont {C.}~\bibnamefont {Nehrbass-Ahles}}, \bibinfo
  {author} {\bibfnamefont {T.~F.}\ \bibnamefont {Stocker}}, \bibinfo {author}
  {\bibfnamefont {H.}~\bibnamefont {Fischer}}, \bibinfo {author} {\bibfnamefont
  {S.}~\bibnamefont {Kipfstuhl}},\ and\ \bibinfo {author} {\bibfnamefont
  {J.}~\bibnamefont {Chappellaz}},\ }\bibfield  {title} {\bibinfo {title}
  {Revision of the epica dome c co2 record from 800 to 600 kyr before
  present},\ }\href@noop {} {\bibfield  {journal} {\bibinfo  {journal}
  {Geophysical Research Letters}\ }\textbf {\bibinfo {volume} {42}},\ \bibinfo
  {pages} {542} (\bibinfo {year} {2015})}\BibitemShut {NoStop}%
\bibitem [{\citenamefont {Loulergue}\ \emph {et~al.}(2008)\citenamefont
  {Loulergue}, \citenamefont {Schilt}, \citenamefont {Spahni}, \citenamefont
  {Masson-Delmotte}, \citenamefont {Blunier}, \citenamefont {Lemieux},
  \citenamefont {Barnola}, \citenamefont {Raynaud}, \citenamefont {Stocker},\
  and\ \citenamefont {Chappellaz}}]{loulergue2008orbital}%
  \BibitemOpen
  \bibfield  {author} {\bibinfo {author} {\bibfnamefont {L.}~\bibnamefont
  {Loulergue}}, \bibinfo {author} {\bibfnamefont {A.}~\bibnamefont {Schilt}},
  \bibinfo {author} {\bibfnamefont {R.}~\bibnamefont {Spahni}}, \bibinfo
  {author} {\bibfnamefont {V.}~\bibnamefont {Masson-Delmotte}}, \bibinfo
  {author} {\bibfnamefont {T.}~\bibnamefont {Blunier}}, \bibinfo {author}
  {\bibfnamefont {B.}~\bibnamefont {Lemieux}}, \bibinfo {author} {\bibfnamefont
  {J.-M.}\ \bibnamefont {Barnola}}, \bibinfo {author} {\bibfnamefont
  {D.}~\bibnamefont {Raynaud}}, \bibinfo {author} {\bibfnamefont {T.~F.}\
  \bibnamefont {Stocker}},\ and\ \bibinfo {author} {\bibfnamefont
  {J.}~\bibnamefont {Chappellaz}},\ }\bibfield  {title} {\bibinfo {title}
  {Orbital and millennial-scale features of atmospheric ch 4 over the past
  800,000 years},\ }\href@noop {} {\bibfield  {journal} {\bibinfo  {journal}
  {Nature}\ }\textbf {\bibinfo {volume} {453}},\ \bibinfo {pages} {383}
  (\bibinfo {year} {2008})}\BibitemShut {NoStop}%
\bibitem [{\citenamefont {Bazin}\ \emph {et~al.}(2013)\citenamefont {Bazin},
  \citenamefont {Landais}, \citenamefont {Lemieux-Dudon}, \citenamefont
  {Toy{\'e} Mahamadou~Kele}, \citenamefont {Veres}, \citenamefont {Parrenin},
  \citenamefont {Martinerie}, \citenamefont {Ritz}, \citenamefont {Capron},
  \citenamefont {Lipenkov} \emph {et~al.}}]{bazin2013optimized}%
  \BibitemOpen
  \bibfield  {author} {\bibinfo {author} {\bibfnamefont {L.}~\bibnamefont
  {Bazin}}, \bibinfo {author} {\bibfnamefont {A.}~\bibnamefont {Landais}},
  \bibinfo {author} {\bibfnamefont {B.}~\bibnamefont {Lemieux-Dudon}}, \bibinfo
  {author} {\bibfnamefont {H.}~\bibnamefont {Toy{\'e} Mahamadou~Kele}},
  \bibinfo {author} {\bibfnamefont {D.}~\bibnamefont {Veres}}, \bibinfo
  {author} {\bibfnamefont {F.}~\bibnamefont {Parrenin}}, \bibinfo {author}
  {\bibfnamefont {P.}~\bibnamefont {Martinerie}}, \bibinfo {author}
  {\bibfnamefont {C.}~\bibnamefont {Ritz}}, \bibinfo {author} {\bibfnamefont
  {E.}~\bibnamefont {Capron}}, \bibinfo {author} {\bibfnamefont
  {V.}~\bibnamefont {Lipenkov}}, \emph {et~al.},\ }\bibfield  {title} {\bibinfo
  {title} {An optimized multi-proxy, multi-site antarctic ice and gas orbital
  chronology (aicc2012): 120--800 ka},\ }\href@noop {} {\bibfield  {journal}
  {\bibinfo  {journal} {Climate of the Past}\ }\textbf {\bibinfo {volume}
  {9}},\ \bibinfo {pages} {1715} (\bibinfo {year} {2013})}\BibitemShut
  {NoStop}%
\bibitem [{\citenamefont {Elderfield}\ \emph {et~al.}(2012)\citenamefont
  {Elderfield}, \citenamefont {Ferretti}, \citenamefont {Greaves},
  \citenamefont {Crowhurst}, \citenamefont {McCave}, \citenamefont {Hodell},\
  and\ \citenamefont {Piotrowski}}]{elderfield2012evolution}%
  \BibitemOpen
  \bibfield  {author} {\bibinfo {author} {\bibfnamefont {H.}~\bibnamefont
  {Elderfield}}, \bibinfo {author} {\bibfnamefont {P.}~\bibnamefont
  {Ferretti}}, \bibinfo {author} {\bibfnamefont {M.}~\bibnamefont {Greaves}},
  \bibinfo {author} {\bibfnamefont {S.}~\bibnamefont {Crowhurst}}, \bibinfo
  {author} {\bibfnamefont {I.~N.}\ \bibnamefont {McCave}}, \bibinfo {author}
  {\bibfnamefont {D.}~\bibnamefont {Hodell}},\ and\ \bibinfo {author}
  {\bibfnamefont {A.~M.}\ \bibnamefont {Piotrowski}},\ }\bibfield  {title}
  {\bibinfo {title} {Evolution of ocean temperature and ice volume through the
  mid-pleistocene climate transition},\ }\href@noop {} {\bibfield  {journal}
  {\bibinfo  {journal} {science}\ }\textbf {\bibinfo {volume} {337}},\ \bibinfo
  {pages} {704} (\bibinfo {year} {2012})}\BibitemShut {NoStop}%
\bibitem [{\citenamefont {Lawrimore}\ \emph {et~al.}(2011)\citenamefont
  {Lawrimore}, \citenamefont {Menne}, \citenamefont {Gleason}, \citenamefont
  {Williams}, \citenamefont {Wuertz}, \citenamefont {Vose},\ and\ \citenamefont
  {Rennie}}]{lawrimore2011overview}%
  \BibitemOpen
  \bibfield  {author} {\bibinfo {author} {\bibfnamefont {J.~H.}\ \bibnamefont
  {Lawrimore}}, \bibinfo {author} {\bibfnamefont {M.~J.}\ \bibnamefont
  {Menne}}, \bibinfo {author} {\bibfnamefont {B.~E.}\ \bibnamefont {Gleason}},
  \bibinfo {author} {\bibfnamefont {C.~N.}\ \bibnamefont {Williams}}, \bibinfo
  {author} {\bibfnamefont {D.~B.}\ \bibnamefont {Wuertz}}, \bibinfo {author}
  {\bibfnamefont {R.~S.}\ \bibnamefont {Vose}},\ and\ \bibinfo {author}
  {\bibfnamefont {J.}~\bibnamefont {Rennie}},\ }\bibfield  {title} {\bibinfo
  {title} {An overview of the global historical climatology network monthly
  mean temperature data set, version 3},\ }\href@noop {} {\bibfield  {journal}
  {\bibinfo  {journal} {Journal of Geophysical Research: Atmospheres}\ }\textbf
  {\bibinfo {volume} {116}} (\bibinfo {year} {2011})}\BibitemShut {NoStop}%
\bibitem [{\citenamefont {Li}\ \emph {et~al.}(2011)\citenamefont {Li},
  \citenamefont {Xie}, \citenamefont {Cook}, \citenamefont {Huang},
  \citenamefont {D'arrigo}, \citenamefont {Liu}, \citenamefont {Ma},\ and\
  \citenamefont {Zheng}}]{li2011interdecadal}%
  \BibitemOpen
  \bibfield  {author} {\bibinfo {author} {\bibfnamefont {J.}~\bibnamefont
  {Li}}, \bibinfo {author} {\bibfnamefont {S.-P.}\ \bibnamefont {Xie}},
  \bibinfo {author} {\bibfnamefont {E.~R.}\ \bibnamefont {Cook}}, \bibinfo
  {author} {\bibfnamefont {G.}~\bibnamefont {Huang}}, \bibinfo {author}
  {\bibfnamefont {R.}~\bibnamefont {D'arrigo}}, \bibinfo {author}
  {\bibfnamefont {F.}~\bibnamefont {Liu}}, \bibinfo {author} {\bibfnamefont
  {J.}~\bibnamefont {Ma}},\ and\ \bibinfo {author} {\bibfnamefont {X.-T.}\
  \bibnamefont {Zheng}},\ }\bibfield  {title} {\bibinfo {title} {Interdecadal
  modulation of el ni{\~n}o amplitude during the past millennium},\ }\href@noop
  {} {\bibfield  {journal} {\bibinfo  {journal} {Nature climate change}\
  }\textbf {\bibinfo {volume} {1}},\ \bibinfo {pages} {114} (\bibinfo {year}
  {2011})}\BibitemShut {NoStop}%
\bibitem [{\citenamefont {Shi}\ \emph {et~al.}(2014)\citenamefont {Shi},
  \citenamefont {Li},\ and\ \citenamefont {Wilson}}]{shi2014tree}%
  \BibitemOpen
  \bibfield  {author} {\bibinfo {author} {\bibfnamefont {F.}~\bibnamefont
  {Shi}}, \bibinfo {author} {\bibfnamefont {J.}~\bibnamefont {Li}},\ and\
  \bibinfo {author} {\bibfnamefont {R.~J.}\ \bibnamefont {Wilson}},\ }\bibfield
   {title} {\bibinfo {title} {A tree-ring reconstruction of the south asian
  summer monsoon index over the past millennium},\ }\href@noop {} {\bibfield
  {journal} {\bibinfo  {journal} {Scientific Reports}\ }\textbf {\bibinfo
  {volume} {4}},\ \bibinfo {pages} {1} (\bibinfo {year} {2014})}\BibitemShut
  {NoStop}%
\bibitem [{\citenamefont {Rayner}\ \emph {et~al.}(2003)\citenamefont {Rayner},
  \citenamefont {Parker}, \citenamefont {Horton}, \citenamefont {Folland},
  \citenamefont {Alexander}, \citenamefont {Rowell}, \citenamefont {Kent},\
  and\ \citenamefont {Kaplan}}]{rayner2003global}%
  \BibitemOpen
  \bibfield  {author} {\bibinfo {author} {\bibfnamefont {N.}~\bibnamefont
  {Rayner}}, \bibinfo {author} {\bibfnamefont {D.~E.}\ \bibnamefont {Parker}},
  \bibinfo {author} {\bibfnamefont {E.}~\bibnamefont {Horton}}, \bibinfo
  {author} {\bibfnamefont {C.~K.}\ \bibnamefont {Folland}}, \bibinfo {author}
  {\bibfnamefont {L.~V.}\ \bibnamefont {Alexander}}, \bibinfo {author}
  {\bibfnamefont {D.}~\bibnamefont {Rowell}}, \bibinfo {author} {\bibfnamefont
  {E.~C.}\ \bibnamefont {Kent}},\ and\ \bibinfo {author} {\bibfnamefont
  {A.}~\bibnamefont {Kaplan}},\ }\bibfield  {title} {\bibinfo {title} {Global
  analyses of sea surface temperature, sea ice, and night marine air
  temperature since the late nineteenth century},\ }\href@noop {} {\bibfield
  {journal} {\bibinfo  {journal} {Journal of Geophysical Research:
  Atmospheres}\ }\textbf {\bibinfo {volume} {108}} (\bibinfo {year}
  {2003})}\BibitemShut {NoStop}%
\bibitem [{\citenamefont {Luterbacher}\ \emph {et~al.}(1999)\citenamefont
  {Luterbacher}, \citenamefont {Schmutz}, \citenamefont {Gyalistras},
  \citenamefont {Xoplaki},\ and\ \citenamefont
  {Wanner}}]{luterbacher1999reconstruction}%
  \BibitemOpen
  \bibfield  {author} {\bibinfo {author} {\bibfnamefont {J.}~\bibnamefont
  {Luterbacher}}, \bibinfo {author} {\bibfnamefont {C.}~\bibnamefont
  {Schmutz}}, \bibinfo {author} {\bibfnamefont {D.}~\bibnamefont {Gyalistras}},
  \bibinfo {author} {\bibfnamefont {E.}~\bibnamefont {Xoplaki}},\ and\ \bibinfo
  {author} {\bibfnamefont {H.}~\bibnamefont {Wanner}},\ }\bibfield  {title}
  {\bibinfo {title} {Reconstruction of monthly nao and eu indices back to ad
  1675},\ }\href@noop {} {\bibfield  {journal} {\bibinfo  {journal}
  {Geophysical Research Letters}\ }\textbf {\bibinfo {volume} {26}},\ \bibinfo
  {pages} {2745} (\bibinfo {year} {1999})}\BibitemShut {NoStop}%
\bibitem [{\citenamefont {Luterbacher}\ \emph {et~al.}(2001)\citenamefont
  {Luterbacher}, \citenamefont {Xoplaki}, \citenamefont {Dietrich},
  \citenamefont {Jones}, \citenamefont {Davies}, \citenamefont {Portis},
  \citenamefont {Gonzalez-Rouco}, \citenamefont {Von~Storch}, \citenamefont
  {Gyalistras}, \citenamefont {Casty} \emph
  {et~al.}}]{luterbacher2001extending}%
  \BibitemOpen
  \bibfield  {author} {\bibinfo {author} {\bibfnamefont {J.}~\bibnamefont
  {Luterbacher}}, \bibinfo {author} {\bibfnamefont {E.}~\bibnamefont
  {Xoplaki}}, \bibinfo {author} {\bibfnamefont {D.}~\bibnamefont {Dietrich}},
  \bibinfo {author} {\bibfnamefont {P.}~\bibnamefont {Jones}}, \bibinfo
  {author} {\bibfnamefont {T.}~\bibnamefont {Davies}}, \bibinfo {author}
  {\bibfnamefont {D.}~\bibnamefont {Portis}}, \bibinfo {author} {\bibfnamefont
  {J.}~\bibnamefont {Gonzalez-Rouco}}, \bibinfo {author} {\bibfnamefont
  {H.}~\bibnamefont {Von~Storch}}, \bibinfo {author} {\bibfnamefont
  {D.}~\bibnamefont {Gyalistras}}, \bibinfo {author} {\bibfnamefont
  {C.}~\bibnamefont {Casty}}, \emph {et~al.},\ }\bibfield  {title} {\bibinfo
  {title} {Extending north atlantic oscillation reconstructions back to 1500},\
  }\href@noop {} {\bibfield  {journal} {\bibinfo  {journal} {Atmospheric
  Science Letters}\ }\textbf {\bibinfo {volume} {2}},\ \bibinfo {pages} {114}
  (\bibinfo {year} {2001})}\BibitemShut {NoStop}%
\bibitem [{\citenamefont {Trenberth}\ and\ \citenamefont
  {Paolino~Jr}(1980)}]{trenberth1980northern}%
  \BibitemOpen
  \bibfield  {author} {\bibinfo {author} {\bibfnamefont {K.~E.}\ \bibnamefont
  {Trenberth}}\ and\ \bibinfo {author} {\bibfnamefont {D.~A.}\ \bibnamefont
  {Paolino~Jr}},\ }\bibfield  {title} {\bibinfo {title} {The northern
  hemisphere sea-level pressure data set: Trends, errors and discontinuities},\
  }\href@noop {} {\bibfield  {journal} {\bibinfo  {journal} {Monthly Weather
  Review}\ }\textbf {\bibinfo {volume} {108}},\ \bibinfo {pages} {855}
  (\bibinfo {year} {1980})}\BibitemShut {NoStop}%
\bibitem [{\citenamefont {Dobrovoln{\`y}}\ \emph {et~al.}(2010)\citenamefont
  {Dobrovoln{\`y}}, \citenamefont {Moberg}, \citenamefont {Br{\'a}zdil},
  \citenamefont {Pfister}, \citenamefont {Glaser}, \citenamefont {Wilson},
  \citenamefont {van Engelen}, \citenamefont {Liman{\'o}wka}, \citenamefont
  {Kiss}, \citenamefont {Hal{\'\i}{\v{c}}kov{\'a}} \emph
  {et~al.}}]{dobrovolny2010monthly}%
  \BibitemOpen
  \bibfield  {author} {\bibinfo {author} {\bibfnamefont {P.}~\bibnamefont
  {Dobrovoln{\`y}}}, \bibinfo {author} {\bibfnamefont {A.}~\bibnamefont
  {Moberg}}, \bibinfo {author} {\bibfnamefont {R.}~\bibnamefont {Br{\'a}zdil}},
  \bibinfo {author} {\bibfnamefont {C.}~\bibnamefont {Pfister}}, \bibinfo
  {author} {\bibfnamefont {R.}~\bibnamefont {Glaser}}, \bibinfo {author}
  {\bibfnamefont {R.}~\bibnamefont {Wilson}}, \bibinfo {author} {\bibfnamefont
  {A.}~\bibnamefont {van Engelen}}, \bibinfo {author} {\bibfnamefont
  {D.}~\bibnamefont {Liman{\'o}wka}}, \bibinfo {author} {\bibfnamefont
  {A.}~\bibnamefont {Kiss}}, \bibinfo {author} {\bibfnamefont {M.}~\bibnamefont
  {Hal{\'\i}{\v{c}}kov{\'a}}}, \emph {et~al.},\ }\bibfield  {title} {\bibinfo
  {title} {Monthly, seasonal and annual temperature reconstructions for central
  europe derived from documentary evidence and instrumental records since ad
  1500},\ }\href@noop {} {\bibfield  {journal} {\bibinfo  {journal} {Climatic
  change}\ }\textbf {\bibinfo {volume} {101}},\ \bibinfo {pages} {69} (\bibinfo
  {year} {2010})}\BibitemShut {NoStop}%
\bibitem [{\citenamefont {Barnston}\ and\ \citenamefont
  {Livezey}(1987)}]{barnston1987classification}%
  \BibitemOpen
  \bibfield  {author} {\bibinfo {author} {\bibfnamefont {A.~G.}\ \bibnamefont
  {Barnston}}\ and\ \bibinfo {author} {\bibfnamefont {R.~E.}\ \bibnamefont
  {Livezey}},\ }\bibfield  {title} {\bibinfo {title} {Classification,
  seasonality and persistence of low-frequency atmospheric circulation
  patterns},\ }\href@noop {} {\bibfield  {journal} {\bibinfo  {journal}
  {Monthly weather review}\ }\textbf {\bibinfo {volume} {115}},\ \bibinfo
  {pages} {1083} (\bibinfo {year} {1987})}\BibitemShut {NoStop}%
\bibitem [{\citenamefont {Chen}\ and\ \citenamefont {Van~den
  Dool}(2003)}]{chen2003sensitivity}%
  \BibitemOpen
  \bibfield  {author} {\bibinfo {author} {\bibfnamefont {W.~Y.}\ \bibnamefont
  {Chen}}\ and\ \bibinfo {author} {\bibfnamefont {H.}~\bibnamefont {Van~den
  Dool}},\ }\bibfield  {title} {\bibinfo {title} {Sensitivity of teleconnection
  patterns to the sign of their primary action center},\ }\href@noop {}
  {\bibfield  {journal} {\bibinfo  {journal} {Monthly weather review}\ }\textbf
  {\bibinfo {volume} {131}},\ \bibinfo {pages} {2885} (\bibinfo {year}
  {2003})}\BibitemShut {NoStop}%
\bibitem [{\citenamefont {Van~den Dool}\ \emph {et~al.}(2000)\citenamefont
  {Van~den Dool}, \citenamefont {Saha},\ and\ \citenamefont
  {Johansson}}]{van2000empirical}%
  \BibitemOpen
  \bibfield  {author} {\bibinfo {author} {\bibfnamefont {H.}~\bibnamefont
  {Van~den Dool}}, \bibinfo {author} {\bibfnamefont {S.}~\bibnamefont {Saha}},\
  and\ \bibinfo {author} {\bibfnamefont {A.}~\bibnamefont {Johansson}},\
  }\bibfield  {title} {\bibinfo {title} {Empirical orthogonal
  teleconnections},\ }\href@noop {} {\bibfield  {journal} {\bibinfo  {journal}
  {Journal of Climate}\ }\textbf {\bibinfo {volume} {13}},\ \bibinfo {pages}
  {1421} (\bibinfo {year} {2000})}\BibitemShut {NoStop}%
\bibitem [{\citenamefont {Klein~Tank}\ \emph {et~al.}(2002)\citenamefont
  {Klein~Tank}, \citenamefont {Wijngaard}, \citenamefont {K{\"o}nnen},
  \citenamefont {B{\"o}hm}, \citenamefont {Demar{\'e}e}, \citenamefont
  {Gocheva}, \citenamefont {Mileta}, \citenamefont {Pashiardis}, \citenamefont
  {Hejkrlik}, \citenamefont {Kern-Hansen} \emph {et~al.}}]{klein2002daily}%
  \BibitemOpen
  \bibfield  {author} {\bibinfo {author} {\bibfnamefont {A.}~\bibnamefont
  {Klein~Tank}}, \bibinfo {author} {\bibfnamefont {J.}~\bibnamefont
  {Wijngaard}}, \bibinfo {author} {\bibfnamefont {G.}~\bibnamefont
  {K{\"o}nnen}}, \bibinfo {author} {\bibfnamefont {R.}~\bibnamefont
  {B{\"o}hm}}, \bibinfo {author} {\bibfnamefont {G.}~\bibnamefont
  {Demar{\'e}e}}, \bibinfo {author} {\bibfnamefont {A.}~\bibnamefont
  {Gocheva}}, \bibinfo {author} {\bibfnamefont {M.}~\bibnamefont {Mileta}},
  \bibinfo {author} {\bibfnamefont {S.}~\bibnamefont {Pashiardis}}, \bibinfo
  {author} {\bibfnamefont {L.}~\bibnamefont {Hejkrlik}}, \bibinfo {author}
  {\bibfnamefont {C.}~\bibnamefont {Kern-Hansen}}, \emph {et~al.},\ }\bibfield
  {title} {\bibinfo {title} {Daily dataset of 20th-century surface air
  temperature and precipitation series for the european climate assessment},\
  }\href@noop {} {\bibfield  {journal} {\bibinfo  {journal} {International
  Journal of Climatology: A Journal of the Royal Meteorological Society}\
  }\textbf {\bibinfo {volume} {22}},\ \bibinfo {pages} {1441} (\bibinfo {year}
  {2002})}\BibitemShut {NoStop}%
\bibitem [{\citenamefont {Politis}\ and\ \citenamefont
  {Romano}(1994)}]{politis1994stationary}%
  \BibitemOpen
  \bibfield  {author} {\bibinfo {author} {\bibfnamefont {D.~N.}\ \bibnamefont
  {Politis}}\ and\ \bibinfo {author} {\bibfnamefont {J.~P.}\ \bibnamefont
  {Romano}},\ }\bibfield  {title} {\bibinfo {title} {The stationary
  bootstrap},\ }\href@noop {} {\bibfield  {journal} {\bibinfo  {journal}
  {Journal of the American Statistical association}\ }\textbf {\bibinfo
  {volume} {89}},\ \bibinfo {pages} {1303} (\bibinfo {year}
  {1994})}\BibitemShut {NoStop}%
\bibitem [{\citenamefont {Foote}(1856)}]{foote1856art}%
  \BibitemOpen
  \bibfield  {author} {\bibinfo {author} {\bibfnamefont {E.}~\bibnamefont
  {Foote}},\ }\bibfield  {title} {\bibinfo {title} {Art. xxxi.--circumstances
  affecting the heat of the sun's rays},\ }\href@noop {} {\bibfield  {journal}
  {\bibinfo  {journal} {American Journal of Science and Arts (1820-1879)}\
  }\textbf {\bibinfo {volume} {22}},\ \bibinfo {pages} {382} (\bibinfo {year}
  {1856})}\BibitemShut {NoStop}%
\bibitem [{\citenamefont {Arrhenius}(1896)}]{arrhenius1896xxxi}%
  \BibitemOpen
  \bibfield  {author} {\bibinfo {author} {\bibfnamefont {S.}~\bibnamefont
  {Arrhenius}},\ }\bibfield  {title} {\bibinfo {title} {Xxxi. on the influence
  of carbonic acid in the air upon the temperature of the ground},\ }\href@noop
  {} {\bibfield  {journal} {\bibinfo  {journal} {The London, Edinburgh, and
  Dublin Philosophical Magazine and Journal of Science}\ }\textbf {\bibinfo
  {volume} {41}},\ \bibinfo {pages} {237} (\bibinfo {year} {1896})}\BibitemShut
  {NoStop}%
\bibitem [{\citenamefont {Kodra}\ \emph {et~al.}(2011)\citenamefont {Kodra},
  \citenamefont {Chatterjee},\ and\ \citenamefont
  {Ganguly}}]{kodra2011exploring}%
  \BibitemOpen
  \bibfield  {author} {\bibinfo {author} {\bibfnamefont {E.}~\bibnamefont
  {Kodra}}, \bibinfo {author} {\bibfnamefont {S.}~\bibnamefont {Chatterjee}},\
  and\ \bibinfo {author} {\bibfnamefont {A.~R.}\ \bibnamefont {Ganguly}},\
  }\bibfield  {title} {\bibinfo {title} {Exploring granger causality between
  global average observed time series of carbon dioxide and temperature},\
  }\href@noop {} {\bibfield  {journal} {\bibinfo  {journal} {Theoretical and
  applied climatology}\ }\textbf {\bibinfo {volume} {104}},\ \bibinfo {pages}
  {325} (\bibinfo {year} {2011})}\BibitemShut {NoStop}%
\bibitem [{\citenamefont {Attanasio}(2012)}]{attanasio2012testing}%
  \BibitemOpen
  \bibfield  {author} {\bibinfo {author} {\bibfnamefont {A.}~\bibnamefont
  {Attanasio}},\ }\bibfield  {title} {\bibinfo {title} {Testing for linear
  granger causality from natural/anthropogenic forcings to global temperature
  anomalies},\ }\href@noop {} {\bibfield  {journal} {\bibinfo  {journal}
  {Theoretical and applied climatology}\ }\textbf {\bibinfo {volume} {110}},\
  \bibinfo {pages} {281} (\bibinfo {year} {2012})}\BibitemShut {NoStop}%
\bibitem [{\citenamefont {Stern}\ and\ \citenamefont
  {Kaufmann}(2014)}]{stern2014anthropogenic}%
  \BibitemOpen
  \bibfield  {author} {\bibinfo {author} {\bibfnamefont {D.~I.}\ \bibnamefont
  {Stern}}\ and\ \bibinfo {author} {\bibfnamefont {R.~K.}\ \bibnamefont
  {Kaufmann}},\ }\bibfield  {title} {\bibinfo {title} {Anthropogenic and
  natural causes of climate change},\ }\href@noop {} {\bibfield  {journal}
  {\bibinfo  {journal} {Climatic change}\ }\textbf {\bibinfo {volume} {122}},\
  \bibinfo {pages} {257} (\bibinfo {year} {2014})}\BibitemShut {NoStop}%
\bibitem [{\citenamefont {Kang}\ and\ \citenamefont
  {Larsson}(2014)}]{kang2014link}%
  \BibitemOpen
  \bibfield  {author} {\bibinfo {author} {\bibfnamefont {J.}~\bibnamefont
  {Kang}}\ and\ \bibinfo {author} {\bibfnamefont {R.}~\bibnamefont {Larsson}},\
  }\bibfield  {title} {\bibinfo {title} {What is the link between temperature
  and carbon dioxide levels? a granger causality analysis based on ice core
  data},\ }\href@noop {} {\bibfield  {journal} {\bibinfo  {journal}
  {Theoretical and applied climatology}\ }\textbf {\bibinfo {volume} {116}},\
  \bibinfo {pages} {537} (\bibinfo {year} {2014})}\BibitemShut {NoStop}%
\bibitem [{\citenamefont {Triacca}(2001)}]{triacca2001use}%
  \BibitemOpen
  \bibfield  {author} {\bibinfo {author} {\bibfnamefont {U.}~\bibnamefont
  {Triacca}},\ }\bibfield  {title} {\bibinfo {title} {On the use of granger
  causality to investigate the human influence on climate},\ }\href@noop {}
  {\bibfield  {journal} {\bibinfo  {journal} {Theoretical and Applied
  Climatology}\ }\textbf {\bibinfo {volume} {69}},\ \bibinfo {pages} {137}
  (\bibinfo {year} {2001})}\BibitemShut {NoStop}%
\bibitem [{\citenamefont {Triacca}(2005)}]{triacca2005granger}%
  \BibitemOpen
  \bibfield  {author} {\bibinfo {author} {\bibfnamefont {U.}~\bibnamefont
  {Triacca}},\ }\bibfield  {title} {\bibinfo {title} {Is granger causality
  analysis appropriate to investigate the relationship between atmospheric
  concentration of carbon dioxide and global surface air temperature?},\
  }\href@noop {} {\bibfield  {journal} {\bibinfo  {journal} {Theoretical and
  applied climatology}\ }\textbf {\bibinfo {volume} {81}},\ \bibinfo {pages}
  {133} (\bibinfo {year} {2005})}\BibitemShut {NoStop}%
\bibitem [{\citenamefont {Stips}\ \emph {et~al.}(2016)\citenamefont {Stips},
  \citenamefont {Macias}, \citenamefont {Coughlan}, \citenamefont
  {Garcia-Gorriz},\ and\ \citenamefont {San~Liang}}]{stips2016causal}%
  \BibitemOpen
  \bibfield  {author} {\bibinfo {author} {\bibfnamefont {A.}~\bibnamefont
  {Stips}}, \bibinfo {author} {\bibfnamefont {D.}~\bibnamefont {Macias}},
  \bibinfo {author} {\bibfnamefont {C.}~\bibnamefont {Coughlan}}, \bibinfo
  {author} {\bibfnamefont {E.}~\bibnamefont {Garcia-Gorriz}},\ and\ \bibinfo
  {author} {\bibfnamefont {X.}~\bibnamefont {San~Liang}},\ }\bibfield  {title}
  {\bibinfo {title} {On the causal structure between co 2 and global
  temperature},\ }\href@noop {} {\bibfield  {journal} {\bibinfo  {journal}
  {Scientific reports}\ }\textbf {\bibinfo {volume} {6}},\ \bibinfo {pages} {1}
  (\bibinfo {year} {2016})}\BibitemShut {NoStop}%
\bibitem [{\citenamefont {Goulet~Coulombe}\ and\ \citenamefont
  {G{\"o}bel}(2021)}]{goulet2021spurious}%
  \BibitemOpen
  \bibfield  {author} {\bibinfo {author} {\bibfnamefont {P.}~\bibnamefont
  {Goulet~Coulombe}}\ and\ \bibinfo {author} {\bibfnamefont {M.}~\bibnamefont
  {G{\"o}bel}},\ }\bibfield  {title} {\bibinfo {title} {On spurious causality,
  co2, and global temperature},\ }\href@noop {} {\bibfield  {journal} {\bibinfo
   {journal} {Econometrics}\ }\textbf {\bibinfo {volume} {9}},\ \bibinfo
  {pages} {33} (\bibinfo {year} {2021})}\BibitemShut {NoStop}%
\bibitem [{\citenamefont {Van~Nes}\ \emph {et~al.}(2015)\citenamefont
  {Van~Nes}, \citenamefont {Scheffer}, \citenamefont {Brovkin}, \citenamefont
  {Lenton}, \citenamefont {Ye}, \citenamefont {Deyle},\ and\ \citenamefont
  {Sugihara}}]{van2015causal}%
  \BibitemOpen
  \bibfield  {author} {\bibinfo {author} {\bibfnamefont {E.~H.}\ \bibnamefont
  {Van~Nes}}, \bibinfo {author} {\bibfnamefont {M.}~\bibnamefont {Scheffer}},
  \bibinfo {author} {\bibfnamefont {V.}~\bibnamefont {Brovkin}}, \bibinfo
  {author} {\bibfnamefont {T.~M.}\ \bibnamefont {Lenton}}, \bibinfo {author}
  {\bibfnamefont {H.}~\bibnamefont {Ye}}, \bibinfo {author} {\bibfnamefont
  {E.}~\bibnamefont {Deyle}},\ and\ \bibinfo {author} {\bibfnamefont
  {G.}~\bibnamefont {Sugihara}},\ }\bibfield  {title} {\bibinfo {title} {Causal
  feedbacks in climate change},\ }\href@noop {} {\bibfield  {journal} {\bibinfo
   {journal} {Nature Climate Change}\ }\textbf {\bibinfo {volume} {5}},\
  \bibinfo {pages} {445} (\bibinfo {year} {2015})}\BibitemShut {NoStop}%
\bibitem [{\citenamefont {Koutsoyiannis}\ and\ \citenamefont
  {Kundzewicz}(2020)}]{koutsoyiannis2020atmospheric}%
  \BibitemOpen
  \bibfield  {author} {\bibinfo {author} {\bibfnamefont {D.}~\bibnamefont
  {Koutsoyiannis}}\ and\ \bibinfo {author} {\bibfnamefont {Z.~W.}\ \bibnamefont
  {Kundzewicz}},\ }\bibfield  {title} {\bibinfo {title} {Atmospheric
  temperature and co2: Hen-or-egg causality?},\ }\href@noop {} {\bibfield
  {journal} {\bibinfo  {journal} {Sci}\ }\textbf {\bibinfo {volume} {2}},\
  \bibinfo {pages} {83} (\bibinfo {year} {2020})}\BibitemShut {NoStop}%
\bibitem [{\citenamefont {M{\o}nster}\ \emph {et~al.}(2017)\citenamefont
  {M{\o}nster}, \citenamefont {Fusaroli}, \citenamefont {Tyl{\'e}n},
  \citenamefont {Roepstorff},\ and\ \citenamefont
  {Sherson}}]{monster2017causal}%
  \BibitemOpen
  \bibfield  {author} {\bibinfo {author} {\bibfnamefont {D.}~\bibnamefont
  {M{\o}nster}}, \bibinfo {author} {\bibfnamefont {R.}~\bibnamefont
  {Fusaroli}}, \bibinfo {author} {\bibfnamefont {K.}~\bibnamefont {Tyl{\'e}n}},
  \bibinfo {author} {\bibfnamefont {A.}~\bibnamefont {Roepstorff}},\ and\
  \bibinfo {author} {\bibfnamefont {J.~F.}\ \bibnamefont {Sherson}},\
  }\bibfield  {title} {\bibinfo {title} {Causal inference from noisy
  time-series data {-} {T}esting the convergent cross-mapping algorithm in the
  presence of noise and external influence},\ }\href@noop {} {\bibfield
  {journal} {\bibinfo  {journal} {Future Generation Computer Systems}\ }\textbf
  {\bibinfo {volume} {73}},\ \bibinfo {pages} {52} (\bibinfo {year}
  {2017})}\BibitemShut {NoStop}%
\bibitem [{\citenamefont {Schiecke}\ \emph {et~al.}(2015)\citenamefont
  {Schiecke}, \citenamefont {Pester}, \citenamefont {Feucht}, \citenamefont
  {Leistritz},\ and\ \citenamefont {Witte}}]{schiecke2015convergent}%
  \BibitemOpen
  \bibfield  {author} {\bibinfo {author} {\bibfnamefont {K.}~\bibnamefont
  {Schiecke}}, \bibinfo {author} {\bibfnamefont {B.}~\bibnamefont {Pester}},
  \bibinfo {author} {\bibfnamefont {M.}~\bibnamefont {Feucht}}, \bibinfo
  {author} {\bibfnamefont {L.}~\bibnamefont {Leistritz}},\ and\ \bibinfo
  {author} {\bibfnamefont {H.}~\bibnamefont {Witte}},\ }\bibfield  {title}
  {\bibinfo {title} {Convergent cross mapping: Basic concept, influence of
  estimation parameters and practical application},\ }in\ \href@noop {} {\emph
  {\bibinfo {booktitle} {2015 37th Annual International Conference of the IEEE
  Engineering in Medicine and Biology Society (EMBC)}}}\ (\bibinfo
  {organization} {IEEE},\ \bibinfo {year} {2015})\ pp.\ \bibinfo {pages}
  {7418--7421}\BibitemShut {NoStop}%
\bibitem [{\citenamefont {Janse}\ \emph {et~al.}(2021)\citenamefont {Janse},
  \citenamefont {Hoekstra}, \citenamefont {Jager}, \citenamefont {Zoccali},
  \citenamefont {Tripepi}, \citenamefont {Dekker},\ and\ \citenamefont {van
  Diepen}}]{janse2021conducting}%
  \BibitemOpen
  \bibfield  {author} {\bibinfo {author} {\bibfnamefont {R.~J.}\ \bibnamefont
  {Janse}}, \bibinfo {author} {\bibfnamefont {T.}~\bibnamefont {Hoekstra}},
  \bibinfo {author} {\bibfnamefont {K.~J.}\ \bibnamefont {Jager}}, \bibinfo
  {author} {\bibfnamefont {C.}~\bibnamefont {Zoccali}}, \bibinfo {author}
  {\bibfnamefont {G.}~\bibnamefont {Tripepi}}, \bibinfo {author} {\bibfnamefont
  {F.~W.}\ \bibnamefont {Dekker}},\ and\ \bibinfo {author} {\bibfnamefont
  {M.}~\bibnamefont {van Diepen}},\ }\bibfield  {title} {\bibinfo {title}
  {Conducting correlation analysis: Important limitations and pitfalls},\
  }\href@noop {} {\bibfield  {journal} {\bibinfo  {journal} {Clinical Kidney
  Journal}\ } (\bibinfo {year} {2021})}\BibitemShut {NoStop}%
\bibitem [{\citenamefont {Brook}\ \emph {et~al.}(1996)\citenamefont {Brook},
  \citenamefont {Sowers},\ and\ \citenamefont {Orchardo}}]{brook1996rapid}%
  \BibitemOpen
  \bibfield  {author} {\bibinfo {author} {\bibfnamefont {E.~J.}\ \bibnamefont
  {Brook}}, \bibinfo {author} {\bibfnamefont {T.}~\bibnamefont {Sowers}},\ and\
  \bibinfo {author} {\bibfnamefont {J.}~\bibnamefont {Orchardo}},\ }\bibfield
  {title} {\bibinfo {title} {Rapid variations in atmospheric methane
  concentration during the past 110,000 years},\ }\href@noop {} {\bibfield
  {journal} {\bibinfo  {journal} {Science}\ }\textbf {\bibinfo {volume}
  {273}},\ \bibinfo {pages} {1087} (\bibinfo {year} {1996})}\BibitemShut
  {NoStop}%
\bibitem [{\citenamefont {Thirumalai}\ \emph {et~al.}(2020)\citenamefont
  {Thirumalai}, \citenamefont {Clemens},\ and\ \citenamefont
  {Partin}}]{thirumalai2020methane}%
  \BibitemOpen
  \bibfield  {author} {\bibinfo {author} {\bibfnamefont {K.}~\bibnamefont
  {Thirumalai}}, \bibinfo {author} {\bibfnamefont {S.~C.}\ \bibnamefont
  {Clemens}},\ and\ \bibinfo {author} {\bibfnamefont {J.~W.}\ \bibnamefont
  {Partin}},\ }\bibfield  {title} {\bibinfo {title} {Methane, monsoons, and
  modulation of millennial-scale climate},\ }\href@noop {} {\bibfield
  {journal} {\bibinfo  {journal} {Geophysical Research Letters}\ }\textbf
  {\bibinfo {volume} {47}},\ \bibinfo {pages} {e2020GL087613} (\bibinfo {year}
  {2020})}\BibitemShut {NoStop}%
\bibitem [{\citenamefont {Kripalani}\ and\ \citenamefont
  {Kulkarni}(1997)}]{kripalani1997rainfall}%
  \BibitemOpen
  \bibfield  {author} {\bibinfo {author} {\bibfnamefont {R.~H.}\ \bibnamefont
  {Kripalani}}\ and\ \bibinfo {author} {\bibfnamefont {A.}~\bibnamefont
  {Kulkarni}},\ }\bibfield  {title} {\bibinfo {title} {Rainfall variability
  over south--east asia—connections with indian monsoon and enso extremes:
  new perspectives},\ }\href@noop {} {\bibfield  {journal} {\bibinfo  {journal}
  {International Journal of Climatology: A Journal of the Royal Meteorological
  Society}\ }\textbf {\bibinfo {volume} {17}},\ \bibinfo {pages} {1155}
  (\bibinfo {year} {1997})}\BibitemShut {NoStop}%
\bibitem [{\citenamefont {Kumar}\ \emph {et~al.}(1999)\citenamefont {Kumar},
  \citenamefont {Rajagopalan},\ and\ \citenamefont
  {Cane}}]{kumar1999weakening}%
  \BibitemOpen
  \bibfield  {author} {\bibinfo {author} {\bibfnamefont {K.~K.}\ \bibnamefont
  {Kumar}}, \bibinfo {author} {\bibfnamefont {B.}~\bibnamefont {Rajagopalan}},\
  and\ \bibinfo {author} {\bibfnamefont {M.~A.}\ \bibnamefont {Cane}},\
  }\bibfield  {title} {\bibinfo {title} {On the weakening relationship between
  the indian monsoon and enso},\ }\href@noop {} {\bibfield  {journal} {\bibinfo
   {journal} {Science}\ }\textbf {\bibinfo {volume} {284}},\ \bibinfo {pages}
  {2156} (\bibinfo {year} {1999})}\BibitemShut {NoStop}%
\bibitem [{\citenamefont {Krishnamurthy}\ and\ \citenamefont
  {Goswami}(2000)}]{krishnamurthy2000indian}%
  \BibitemOpen
  \bibfield  {author} {\bibinfo {author} {\bibfnamefont {V.}~\bibnamefont
  {Krishnamurthy}}\ and\ \bibinfo {author} {\bibfnamefont {B.~N.}\ \bibnamefont
  {Goswami}},\ }\bibfield  {title} {\bibinfo {title} {Indian monsoon--enso
  relationship on interdecadal timescale},\ }\href@noop {} {\bibfield
  {journal} {\bibinfo  {journal} {Journal of climate}\ }\textbf {\bibinfo
  {volume} {13}},\ \bibinfo {pages} {579} (\bibinfo {year} {2000})}\BibitemShut
  {NoStop}%
\bibitem [{\citenamefont {Sarkar}\ \emph {et~al.}(2004)\citenamefont {Sarkar},
  \citenamefont {Singh},\ and\ \citenamefont {Kafatos}}]{sarkar2004further}%
  \BibitemOpen
  \bibfield  {author} {\bibinfo {author} {\bibfnamefont {S.}~\bibnamefont
  {Sarkar}}, \bibinfo {author} {\bibfnamefont {R.~P.}\ \bibnamefont {Singh}},\
  and\ \bibinfo {author} {\bibfnamefont {M.}~\bibnamefont {Kafatos}},\
  }\bibfield  {title} {\bibinfo {title} {Further evidences for the weakening
  relationship of indian rainfall and enso over india},\ }\href@noop {}
  {\bibfield  {journal} {\bibinfo  {journal} {Geophysical research letters}\
  }\textbf {\bibinfo {volume} {31}} (\bibinfo {year} {2004})}\BibitemShut
  {NoStop}%
\bibitem [{\citenamefont {Maraun}\ and\ \citenamefont
  {Kurths}(2005)}]{maraun2005epochs}%
  \BibitemOpen
  \bibfield  {author} {\bibinfo {author} {\bibfnamefont {D.}~\bibnamefont
  {Maraun}}\ and\ \bibinfo {author} {\bibfnamefont {J.}~\bibnamefont
  {Kurths}},\ }\bibfield  {title} {\bibinfo {title} {Epochs of phase coherence
  between el nino/southern oscillation and indian monsoon},\ }\href@noop {}
  {\bibfield  {journal} {\bibinfo  {journal} {Geophysical Research Letters}\
  }\textbf {\bibinfo {volume} {32}} (\bibinfo {year} {2005})}\BibitemShut
  {NoStop}%
\bibitem [{\citenamefont {Zubair}\ and\ \citenamefont
  {Ropelewski}(2006)}]{zubair2006strengthening}%
  \BibitemOpen
  \bibfield  {author} {\bibinfo {author} {\bibfnamefont {L.}~\bibnamefont
  {Zubair}}\ and\ \bibinfo {author} {\bibfnamefont {C.~F.}\ \bibnamefont
  {Ropelewski}},\ }\bibfield  {title} {\bibinfo {title} {The strengthening
  relationship between enso and northeast monsoon rainfall over sri lanka and
  southern india},\ }\href@noop {} {\bibfield  {journal} {\bibinfo  {journal}
  {Journal of Climate}\ }\textbf {\bibinfo {volume} {19}},\ \bibinfo {pages}
  {1567} (\bibinfo {year} {2006})}\BibitemShut {NoStop}%
\bibitem [{\citenamefont {Mokhov}\ \emph {et~al.}(2011)\citenamefont {Mokhov},
  \citenamefont {Smirnov}, \citenamefont {Nakonechny}, \citenamefont
  {Kozlenko}, \citenamefont {Seleznev},\ and\ \citenamefont
  {Kurths}}]{mokhov2011alternating}%
  \BibitemOpen
  \bibfield  {author} {\bibinfo {author} {\bibfnamefont {I.~I.}\ \bibnamefont
  {Mokhov}}, \bibinfo {author} {\bibfnamefont {D.~A.}\ \bibnamefont {Smirnov}},
  \bibinfo {author} {\bibfnamefont {P.~I.}\ \bibnamefont {Nakonechny}},
  \bibinfo {author} {\bibfnamefont {S.~S.}\ \bibnamefont {Kozlenko}}, \bibinfo
  {author} {\bibfnamefont {E.~P.}\ \bibnamefont {Seleznev}},\ and\ \bibinfo
  {author} {\bibfnamefont {J.}~\bibnamefont {Kurths}},\ }\bibfield  {title}
  {\bibinfo {title} {Alternating mutual influence of el-ni{\~n}o/southern
  oscillation and indian monsoon},\ }\href@noop {} {\bibfield  {journal}
  {\bibinfo  {journal} {Geophysical Research Letters}\ }\textbf {\bibinfo
  {volume} {38}} (\bibinfo {year} {2011})}\BibitemShut {NoStop}%
\bibitem [{\citenamefont {Mokhov}\ \emph {et~al.}(2012)\citenamefont {Mokhov},
  \citenamefont {Smirnov}, \citenamefont {Nakonechny}, \citenamefont
  {Kozlenko},\ and\ \citenamefont {Kurths}}]{mokhov2012relationship}%
  \BibitemOpen
  \bibfield  {author} {\bibinfo {author} {\bibfnamefont {I.}~\bibnamefont
  {Mokhov}}, \bibinfo {author} {\bibfnamefont {D.}~\bibnamefont {Smirnov}},
  \bibinfo {author} {\bibfnamefont {P.}~\bibnamefont {Nakonechny}}, \bibinfo
  {author} {\bibfnamefont {S.}~\bibnamefont {Kozlenko}},\ and\ \bibinfo
  {author} {\bibfnamefont {J.}~\bibnamefont {Kurths}},\ }\bibfield  {title}
  {\bibinfo {title} {Relationship between el-nino/southern oscillation and the
  indian monsoon},\ }\href@noop {} {\bibfield  {journal} {\bibinfo  {journal}
  {Izvestiya, Atmospheric and Oceanic Physics}\ }\textbf {\bibinfo {volume}
  {48}},\ \bibinfo {pages} {47} (\bibinfo {year} {2012})}\BibitemShut {NoStop}%
\bibitem [{\citenamefont {Le}\ \emph {et~al.}(2020)\citenamefont {Le},
  \citenamefont {Ha}, \citenamefont {Bae},\ and\ \citenamefont
  {Kim}}]{le2020causal}%
  \BibitemOpen
  \bibfield  {author} {\bibinfo {author} {\bibfnamefont {T.}~\bibnamefont
  {Le}}, \bibinfo {author} {\bibfnamefont {K.-J.}\ \bibnamefont {Ha}}, \bibinfo
  {author} {\bibfnamefont {D.-H.}\ \bibnamefont {Bae}},\ and\ \bibinfo {author}
  {\bibfnamefont {S.-H.}\ \bibnamefont {Kim}},\ }\bibfield  {title} {\bibinfo
  {title} {Causal effects of indian ocean dipole on el ni{\~n}o--southern
  oscillation during 1950--2014 based on high-resolution models and reanalysis
  data},\ }\href@noop {} {\bibfield  {journal} {\bibinfo  {journal}
  {Environmental Research Letters}\ }\textbf {\bibinfo {volume} {15}},\
  \bibinfo {pages} {1040b6} (\bibinfo {year} {2020})}\BibitemShut {NoStop}%
\bibitem [{\citenamefont {Wanner}\ \emph {et~al.}(2001)\citenamefont {Wanner},
  \citenamefont {Br{\"o}nnimann}, \citenamefont {Casty}, \citenamefont
  {Gyalistras}, \citenamefont {Luterbacher}, \citenamefont {Schmutz},
  \citenamefont {Stephenson},\ and\ \citenamefont {Xoplaki}}]{wanner2001north}%
  \BibitemOpen
  \bibfield  {author} {\bibinfo {author} {\bibfnamefont {H.}~\bibnamefont
  {Wanner}}, \bibinfo {author} {\bibfnamefont {S.}~\bibnamefont
  {Br{\"o}nnimann}}, \bibinfo {author} {\bibfnamefont {C.}~\bibnamefont
  {Casty}}, \bibinfo {author} {\bibfnamefont {D.}~\bibnamefont {Gyalistras}},
  \bibinfo {author} {\bibfnamefont {J.}~\bibnamefont {Luterbacher}}, \bibinfo
  {author} {\bibfnamefont {C.}~\bibnamefont {Schmutz}}, \bibinfo {author}
  {\bibfnamefont {D.~B.}\ \bibnamefont {Stephenson}},\ and\ \bibinfo {author}
  {\bibfnamefont {E.}~\bibnamefont {Xoplaki}},\ }\bibfield  {title} {\bibinfo
  {title} {North atlantic oscillation--concepts and studies},\ }\href@noop {}
  {\bibfield  {journal} {\bibinfo  {journal} {Surveys in geophysics}\ }\textbf
  {\bibinfo {volume} {22}},\ \bibinfo {pages} {321} (\bibinfo {year}
  {2001})}\BibitemShut {NoStop}%
\bibitem [{\citenamefont {Hurrell}\ and\ \citenamefont
  {Deser}(2010)}]{hurrell2010north}%
  \BibitemOpen
  \bibfield  {author} {\bibinfo {author} {\bibfnamefont {J.~W.}\ \bibnamefont
  {Hurrell}}\ and\ \bibinfo {author} {\bibfnamefont {C.}~\bibnamefont
  {Deser}},\ }\bibfield  {title} {\bibinfo {title} {North atlantic climate
  variability: the role of the north atlantic oscillation},\ }\href@noop {}
  {\bibfield  {journal} {\bibinfo  {journal} {Journal of marine systems}\
  }\textbf {\bibinfo {volume} {79}},\ \bibinfo {pages} {231} (\bibinfo {year}
  {2010})}\BibitemShut {NoStop}%
\bibitem [{\citenamefont {Deser}\ \emph {et~al.}(2017)\citenamefont {Deser},
  \citenamefont {Hurrell},\ and\ \citenamefont {Phillips}}]{deser2017role}%
  \BibitemOpen
  \bibfield  {author} {\bibinfo {author} {\bibfnamefont {C.}~\bibnamefont
  {Deser}}, \bibinfo {author} {\bibfnamefont {J.~W.}\ \bibnamefont {Hurrell}},\
  and\ \bibinfo {author} {\bibfnamefont {A.~S.}\ \bibnamefont {Phillips}},\
  }\bibfield  {title} {\bibinfo {title} {The role of the north atlantic
  oscillation in european climate projections},\ }\href@noop {} {\bibfield
  {journal} {\bibinfo  {journal} {Climate dynamics}\ }\textbf {\bibinfo
  {volume} {49}},\ \bibinfo {pages} {3141} (\bibinfo {year}
  {2017})}\BibitemShut {NoStop}%
\bibitem [{\citenamefont {Wang}\ \emph {et~al.}(2004)\citenamefont {Wang},
  \citenamefont {Anderson}, \citenamefont {Kaufmann},\ and\ \citenamefont
  {Myneni}}]{wang2004relation}%
  \BibitemOpen
  \bibfield  {author} {\bibinfo {author} {\bibfnamefont {W.}~\bibnamefont
  {Wang}}, \bibinfo {author} {\bibfnamefont {B.~T.}\ \bibnamefont {Anderson}},
  \bibinfo {author} {\bibfnamefont {R.~K.}\ \bibnamefont {Kaufmann}},\ and\
  \bibinfo {author} {\bibfnamefont {R.~B.}\ \bibnamefont {Myneni}},\ }\bibfield
   {title} {\bibinfo {title} {The relation between the north atlantic
  oscillation and ssts in the north atlantic basin},\ }\href@noop {} {\bibfield
   {journal} {\bibinfo  {journal} {Journal of Climate}\ }\textbf {\bibinfo
  {volume} {17}},\ \bibinfo {pages} {4752} (\bibinfo {year}
  {2004})}\BibitemShut {NoStop}%
\bibitem [{\citenamefont {Wang}\ \emph {et~al.}(2019)\citenamefont {Wang},
  \citenamefont {Zhang}, \citenamefont {Fan},\ and\ \citenamefont
  {Palus}}]{wang2019central}%
  \BibitemOpen
  \bibfield  {author} {\bibinfo {author} {\bibfnamefont {G.}~\bibnamefont
  {Wang}}, \bibinfo {author} {\bibfnamefont {N.}~\bibnamefont {Zhang}},
  \bibinfo {author} {\bibfnamefont {K.}~\bibnamefont {Fan}},\ and\ \bibinfo
  {author} {\bibfnamefont {M.}~\bibnamefont {Palus}},\ }\bibfield  {title}
  {\bibinfo {title} {Central european air temperature: driving force analysis
  and causal influence of nao},\ }\href@noop {} {\bibfield  {journal} {\bibinfo
   {journal} {Theoretical and Applied Climatology}\ }\textbf {\bibinfo {volume}
  {137}},\ \bibinfo {pages} {1421} (\bibinfo {year} {2019})}\BibitemShut
  {NoStop}%
\bibitem [{\citenamefont {Hlinka}\ \emph {et~al.}(2017)\citenamefont {Hlinka},
  \citenamefont {Jajcay}, \citenamefont {Hartman},\ and\ \citenamefont
  {Palu{\v{s}}}}]{hlinka2017smooth}%
  \BibitemOpen
  \bibfield  {author} {\bibinfo {author} {\bibfnamefont {J.}~\bibnamefont
  {Hlinka}}, \bibinfo {author} {\bibfnamefont {N.}~\bibnamefont {Jajcay}},
  \bibinfo {author} {\bibfnamefont {D.}~\bibnamefont {Hartman}},\ and\ \bibinfo
  {author} {\bibfnamefont {M.}~\bibnamefont {Palu{\v{s}}}},\ }\bibfield
  {title} {\bibinfo {title} {Smooth information flow in temperature climate
  network reflects mass transport},\ }\href@noop {} {\bibfield  {journal}
  {\bibinfo  {journal} {Chaos: An Interdisciplinary Journal of Nonlinear
  Science}\ }\textbf {\bibinfo {volume} {27}},\ \bibinfo {pages} {035811}
  (\bibinfo {year} {2017})}\BibitemShut {NoStop}%
\bibitem [{\citenamefont {Nagaraj}\ \emph {et~al.}(2013)\citenamefont
  {Nagaraj}, \citenamefont {Balasubramanian},\ and\ \citenamefont
  {Dey}}]{nagaraj2013new}%
  \BibitemOpen
  \bibfield  {author} {\bibinfo {author} {\bibfnamefont {N.}~\bibnamefont
  {Nagaraj}}, \bibinfo {author} {\bibfnamefont {K.}~\bibnamefont
  {Balasubramanian}},\ and\ \bibinfo {author} {\bibfnamefont {S.}~\bibnamefont
  {Dey}},\ }\bibfield  {title} {\bibinfo {title} {A new complexity measure for
  time series analysis and classification},\ }\href@noop {} {\bibfield
  {journal} {\bibinfo  {journal} {The European Physical Journal Special
  Topics}\ }\textbf {\bibinfo {volume} {222}},\ \bibinfo {pages} {847}
  (\bibinfo {year} {2013})}\BibitemShut {NoStop}%
\end{thebibliography}%

\end{document}